\documentclass{nature-pre}
\usepackage{times}
\usepackage{amsmath}
\usepackage{amssymb}
\usepackage{comment}
\usepackage{multirow}
\usepackage{physics}
\usepackage{makecell}

\usepackage[utf8]{inputenc}

\usepackage{siunitx}

\usepackage{graphicx}
\graphicspath{./}
\usepackage{placeins}
\usepackage{threeparttable}
\usepackage{float}
\usepackage{placeins}
\usepackage{color, colortbl}
\definecolor{lightgray}{gray}{0.9}

\usepackage{hyperref}

\usepackage{xr-hyper}
\hypersetup{
 bookmarks=true,		% show bookmarks bar?
 unicode=false,			% nonLatin characters in Acrobat’s bookmarks
 pdftoolbar=true,		% show Acrobat’s toolbar?
 pdffitwindow=false,		% window fit to page when opened
 pdfstartview={FitH},		% fits the width of the page to the windo\textbf{\textbf{}}w
 pdfcreator={pdflatex},		% creator of the document
 pdfnewwindow=true,		% links in new window
 colorlinks=true,		% false: boxed links; true: colored links
 linktoc=page,			% defines which part of an entry in the table of contents is made into a link
 linkcolor=blue,		% color of internal links	(red)
 citecolor=blue,		% color of links to bibliography
 filecolor=blue,		% color of file links
 urlcolor=blue			% color of external links
}

\begin{document}

\title{Natural van der Waals silicates as hosts for telecom quantum emitters: the case of erbium-doped talc}

\author{Gell\'{e}rt Dolecsek$^{1,2}$, Zsolt Benedek$^{1,2}$, Nguyen Tien Son$^{3}$, Viktor Iv\'{a}dy$^{1,2,*}$}

\maketitle

\begin{affiliations}
\item{Department of Physics of Complex Systems, Eötvös Loránd University, Egyetem tér 1-3, H-1053 Budapest, Hungary}
\item {MTA–ELTE Lend\"{u}let "Momentum" NewQubit Research Group, Pázmány Péter, Sétány 1/A, 1117 Budapest, Hungary}
\item {Department of Physics, Chemistry and Biology, Linköping University, 58183 Linköping, Sweden}
\item[*] email: ivady.viktor@ttk.elte.hu
\end{affiliations}

\date{\today}                       

\vspace{1cm}

\begin{abstract}
Erbium ion is among the most promising solid-state single photon emitters and spin-photon interfaces for quantum networks, emitting directly in the telecom C-band in many host semiconductors. Recently, the search for scalable, low-noise host materials turned toward atomically thin and van der Waals materials that enable efficient integration with nanophotonic architectures. Here, we identify talc, a naturally occurring layered magnesium silicate, as a promising host for telecom-active erbium centers. Using first-principles density functional theory combined with multireference wavefunction calculations, we investigate the thermodynamic stability, electronic structure, crystal-field splitting, and optical transitions of erbium-related defects in talc. We find that substitutional incorporation of Er at Mg sites is energetically favourable over a wide range of Fermi-levels, leading predominantly to telecom C band emitting Er$^{3+}$ configuration. The characteristic ${^4}I_{13/2} \rightarrow {^4}I_{15/2}$ transition of Er$^{3+}$ is preserved in the talc environment and remains centred near 1.55~$\mu$m, while crystal-field interactions produce a Stark manifold suitable for spectrally selective optical addressing. The combination of thermodynamic stability, wide band gap, low background emission, and compatibility with van der Waals heterostructures suggests that erbium-doped talc constitutes a promising platform for integrated photonics in the C-band.

\end{abstract}

\newpage

\section*{Introduction}

Rare earth elements (REEs), which consist of scandium, yttrium, and the 15 lanthanides\cite{balaram2019rare}, are most commonly present in the environment in the form of $\mathrm{REE^{3+}}$ ions\cite{bunzli2005taking,eliseeva2010lanthanide}. In these chemical structures, the 4f electronic subshell is partially filled, and its electrons are strongly shielded by the outer 5s and 5p subshells\cite{bunzli2005taking,eliseeva2010lanthanide}. Therefore, 4f electrons hardly participate in chemical bonding and are only weakly affected by the surrounding environment\cite{bunzli2005taking}. 

The intra-4f transitions of rare-earth ions are widely exploited in photonic and optoelectronic technologies\cite{bunzli2005taking,eliseeva2010lanthanide}. In particular, rare-earth-doped materials serve as active media in solid-state lasers, optical amplifiers, and infrared-to-visible upconversion for lighting and displays\cite{snitzer1961optical,mears1987low_noise,eliseeva2010lanthanide}. Their long-lived and spectrally sharp luminescence is also used in optical sensing, time-resolved bioimaging, and   security marking devices\cite{bunzli2010biomedical,auzel2004upconversion,eliseeva2010lanthanide}.

In recent years, potential quantum applications have also been demonstrated. Single-photon emission from individual rare-earth ions has been experimentally observed\cite{dibos2018,zhong2018optically,yang2023,ourari2023indistinguishable,yu2023frequency} and has already been used to demonstrate single-ion spin-photon entanglement\cite{uysal2025spin_photon}. For the time being, active research is conducted targeting rare-earth-ion quantum repeaters, scalable networks, and practical rare-earth single-photon sources\cite{kimiaeeasadi2018quantum,ortu2022storage,ruskuc2025multiplexed,yang2023}.

Among the 4f-related optical transitions of REEs, the photoemission of the erbium cation (Er$^{3+}$) deserves particular interest: the characteristic zero-phonon line (ZPL) of Er$^{3+}$ appears near 1550 nm in many hosts\cite{mears1987low_noise,miniscalco1991erbium,ourari2023indistinguishable,bader_analysis_2024,garcia_arellano2025erbium_implanted,seth2025spin_decoherence,fruh_spectral_2026} lying within the telecommunication wavelength range of 1260-1675~nm (0.74-0.98 eV). Moreover, indistinguishable single photon emission\cite{ourari2023indistinguishable,fruh_spectral_2026},  optically detected magnetism resonance (ODMR), spin-photon interface, single-shot readout, and 23~ms long coherence time of Er$^{3+}$ spin qubits have been demonstrated\cite{le_dantec_twenty-threemillisecond_2021,Gritsch2025ErSingleShot,lyasota_narrow_2026}. Furthermore, erbium-doped optical fibers were suggested to be applicable as quantum memory and light storage platforms \cite{gohari_kamel_erbium-doped_2026}.

Recently, two-dimensional and van der Waals materials have attracted considerable interest because their atomically thin geometry enables unusual electronic, optical, spin, and valley properties, as well as assembly into layered heterostructures\cite{Novoselov_2005,wang2012electronics,geim2013van_der_waals,schaibley2016valleytronics,glavin2020emerging}. Their accessible surfaces and efficient coupling to nearby photonic structures also make them promising hosts for optically active defects and quantum emitters\cite{turunen2022quantum,parto2022}. In the case of Er$^{3+}$ ions as dopants, the incorporation into a 2D host has already been demonstrated in WS$_2$, producing telecom-range emission and illustrating the potential of rare-earth-doped van der Waals materials\cite{garcia_arellano2025erbium_implanted}.

Graphene and hexagonal boron nitride are currently among the most prominent deliberately prepared or mechanically exfoliated 2D materials in materials and quantum science\cite{Novoselov_2005,tran2016quantum,gottscholl2020initialization}. However, layered van der Waals structures also occur naturally in abundant silicates, including muscovite, biotite, phlogopite, lepidolite, and talc, which can be exfoliated into thin flakes\cite{frisenda2020naturally,dolecsek2025native}. These minerals therefore, provide a complementary route toward chemically stable 2D platforms without requiring epitaxial or bottom-up synthesis.

Motivated by a previous theoretical study of our group~\cite{dolecsek2025native}, we herein suggest talc as a promising alternative host for Er$^{3+}$ centers. Talc is a wide-band-gap magnesium silicate with a low concentration of optically active intrinsic defects, which may suppress host absorption,  optical background, and charge fluctuation, potentially enabling lifetime-limited line widths for emitters.\cite{dolecsek2025native}. Its atomically flat, dangling-bond-free surfaces and exfoliability\cite{frisenda2020naturally} facilitate the integration and controlled positioning of optically active centers in van der Waals heterostructures and hybrid photonic devices\cite{alencar2015talc,harvey2017exploring,barcelos2018infrared}. In particular, an Er-doped talc flake could be placed near the optical-field maximum of an external dielectric or photonic-crystal cavity, making it suitable for Purcell-enhanced emission despite the moderate refractive index of talc itself\cite{vahala2003optical,parto2022}. Furthermore, being the softest crystal (1 on the Mohs Hardness Scale), talc facilitates the creation of high-quality factor cavities and potentially enables talc-based quantum optics. 

Despite their apparent potential, very little is known about the REEs in talc.
In the present study, we apply density functional theory (DFT) and wavefunction theory (WFT) calculations to characterize erbium structures embedded in talc. After assessing the conceivable configurations and charge states of Er in talc, we investigate their formation energy, electronic structure, optical properties, and the effect of the talc environment on the telecom-band photoemission. Our results demonstrate the practical applicability of hosting Er$^{3+}$ in talc, which opens the path for developing novel quantum optics and nano photonic applications in talc quasi-2D layers.

\begin{figure}[h]
\includegraphics[width=\linewidth]{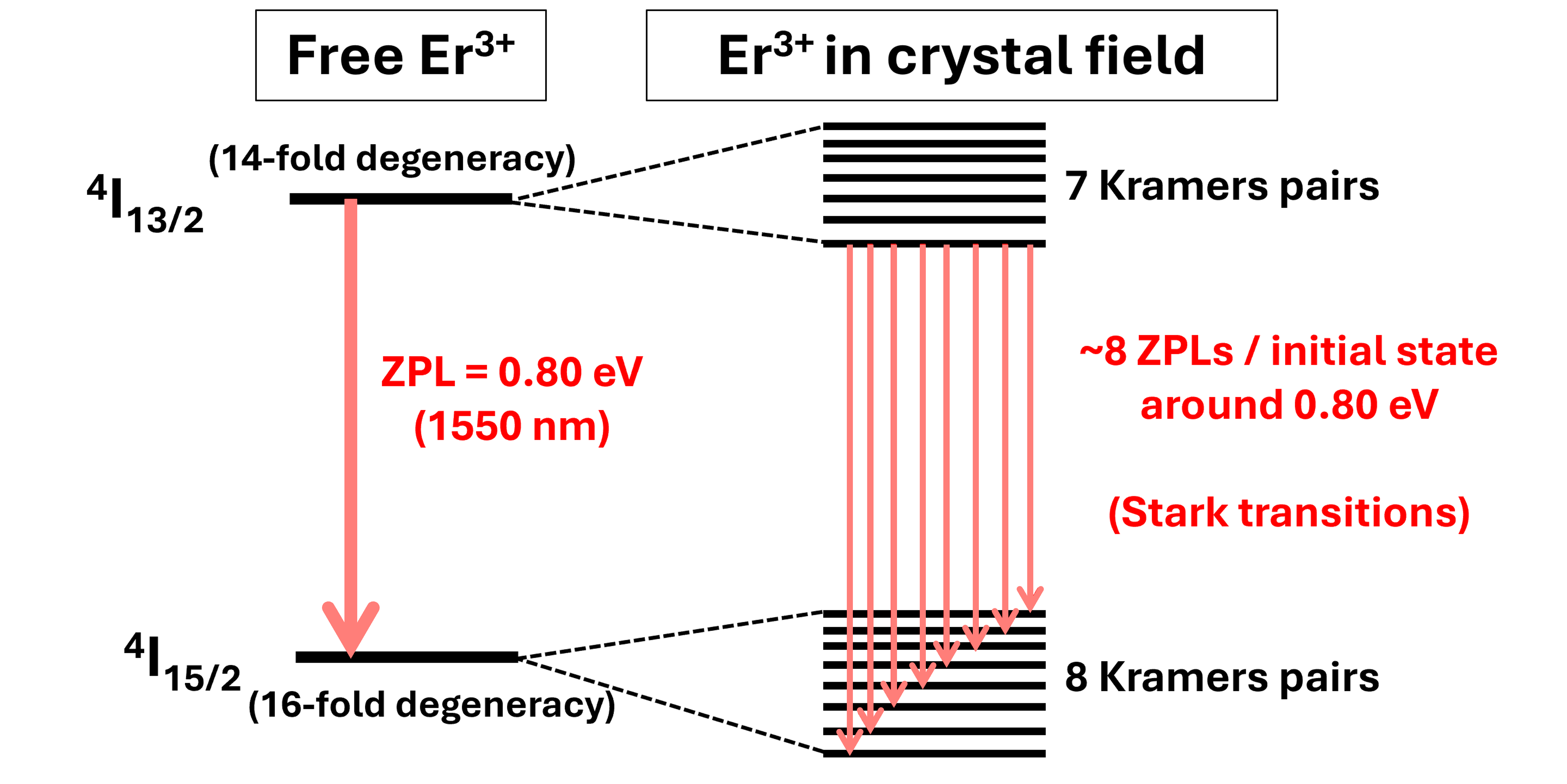}
	\caption{Schematic representation of the lowest-lying spin-orbit coupled energy levels of the free Er$^{3+}$ ion (left) and a crystal-embedded Er$^{3+}$ ion (right).}
	\label{fig:Er}  
\end{figure}

\subsection{Background.}

The well-characterized 1550~nm emission of the Er$^{3+}$ ion is not related to any HOMO--LUMO-type excitation. Instead, the optical transition occurs within the $4f^{11}$ configuration, between two spin orbit-coupled manifolds\cite{carnall1989systematic,bunzli2010basics,miniscalco1991erbium}. The free Er$^{3+}$ ion has 11 electrons (in other words, 3 holes) distributed over the 7 orbitals of the 4f subshell. According to Hund's first, second and third rules, the ground state is $^4I_{15/2}$ with a quartet spin $((2S+1)=4)$ due to three unpaired electrons, with total orbital angular momentum $L=6$ (corresponding to the $I$ term, which can be viewed, in the hole picture, as arising from the maximally allowed individual $m_l$ quantum numbers of 3, 2 and 1) and total angular momentum $J=15/2$, which is the highest possible quantum number within the $(L-S),\ldots,(L+S)$ range for the more-than-half-filled $4f^{11}$ shell. The lowest-lying excited manifold, $^4I_{13/2}$, possesses the same $S$ and $L$ quantum numbers, but a decreased $J$ of 13/2. Therefore, the origin of the telecom band emission is the $^4I_{13/2}\rightarrow{}^4I_{15/2}$ relaxation\cite{carnall1989systematic,desurvire1990evaluation,miniscalco1991erbium}. For such a transition, the photoluminescence spectrum is expected to contain a single sharp ZPL, with negligible phonon sideband due to the absence of geometry relaxation effects.

In free Er$^{3+}$ ions, the aforementioned $(2J+1)$-fold degeneracy is exact, apart from hyperfine interactions and external fields (Fig.~\ref{fig:Er}, left). In a crystal field, however, states belonging to the same $J$ manifold lose their degeneracy. Since Er$^{3+}$ is a Kramers ion with an odd number of 4f electrons, each Stark level remains at least doubly degenerate, and a low-symmetry crystal field can split a half-integer-$J$ manifold into up to $(J+1/2)$ Kramers pairs (Fig.~\ref{fig:Er}, right)\cite{bunzli2010basics,carnall1989systematic}. Consequently, multiple zero-phonon lines are expected in the photoluminescence spectrum, corresponding to transitions between Stark components of the $^4I_{13/2}$ and $^4I_{15/2}$ manifolds\cite{desurvire1990evaluation,miniscalco1991erbium}.
%For example, $(J+1/2)$ gives 8 Kramers pairs for the $^4I_{15/2}$ ground manifold of Er$^{3+}$;
The actual number of observed lines also depends on site symmetry and experimental conditions, such as spectral resolution and temperature\cite{desurvire1990evaluation}. Such splitting does not prevent conventional optoelectronic use of Er$^{3+}$ emission; in fact, it contributes to the spectral structure and bandwidth of Er-doped optical amplifiers. Quantum applications, however, require a well-defined narrow emission line, i.e., one selected Stark transition. To date, reported Er$^{3+}$-based single-ion single-photon sources have relied on Purcell enhancement of selected optical transitions\cite{dibos2018,yang2023,yu2023frequency,ourari2023indistinguishable}.

%\textit{Incorporation of Er$^{3+}$ into solid-state hosts.}

Er$^{3+}$ has been incorporated into diverse solid-state hosts, most prominently silica glasses and optical fibers, which offer chemical and thermal stability, telecom transparency, low optical loss, and mature processing technologies\cite{mears1987low_noise,miniscalco1991erbium}. Common crystalline hosts include Y$_2$SiO$_5$, Y$_2$O$_3$, YVO$_4$, CaWO$_4$, and LiNbO$_3$, selected for their ordered and stable rare-earth coordination environments, low disorder, and good optical transparency\cite{thiel2011rare,bottger2006spectroscopy,gupta2025dual,xie2021characterization,ourari2023indistinguishable}. LiNbO$_3$ additionally provides strong electro-optic response and established waveguide-fabrication routes\cite{zhu2021integrated,wang2020incorporation}. Er$^{3+}$ has also been incorporated into TiO$_2$, CeO$_2$, and other epitaxial or thin-film oxides suitable for thin-film growth and nanophotonic integration\cite{dibos2022purcell,zhang2024optical_spin_ceo2,gupta2025dual}. Less conventional semiconductor hosts, including silicon and SiC, provide compatibility with semiconductor processing, although their Er centers are strongly influenced by local defects, charge compensation, and coordination\cite{yin2013optical,weiss2021erbium,gritsch2022narrow,prezzi2004optical,parker2021infrared}.

Quantum-optical applications have been pursued prominently in Er$^{3+}$:Y$_2$SiO$_5$ for ensemble quantum memories\cite{ortu2022storage,liu2022ondemand}, in Er$^{3+}$:CaWO$_4$ for single-ion spin-photon interfaces and indistinguishable-photon generation\cite{ourari2023indistinguishable,uysal2025spin_photon}, and in LiNbO$_3$ and silicon for integrated single-ion and cavity-coupled photonics\cite{wang2020incorporation,yang2023,yin2013optical,dibos2018,weiss2021erbium,gritsch2022narrow}. Emerging quantum-oriented studies have also considered epitaxial oxide hosts and SiC\cite{dibos2022purcell,zhang2024optical_spin_ceo2,gupta2025dual,parker2021infrared}.

In all these hosts, the relevant telecom photoluminescence is assigned to the shielded intra-4f $^4I_{13/2}\rightarrow{}^4I_{15/2}$ transition of Er$^{3+}$ near 1.5-1.6~$\mu$m, however, the exact wavelength, Stark structure, linewidth, and lifetime are host-dependent\cite{carnall1989systematic,miniscalco1991erbium,bottger2006spectroscopy}. Representative bulk or weakly coupled excited-state lifetimes, referring to the total decay of the $^4I_{13/2}$ manifold through all allowed Stark-resolved emission channels rather than to a single zero-phonon line, are approximately 10~ms in silica fibers\cite{miniscalco1991erbium}, 11.4 and 9.2~ms for the two Er$^{3+}$ sites in Y$_2$SiO$_5$\cite{bottger2006spectroscopy}, and a few to several milliseconds in CaWO$_4$, with reported values depending on site, sample preparation, temperature, and concentration\cite{cornacchia2007growth,becker2025spectroscopic}. In LiNbO$_3$, uncoupled Er$^{3+}$ centers generally exhibit millisecond-scale total excited-state lifetimes, typically of order 2-6~ms depending on the material and processing route\cite{fleuster1994optical,wang2020incorporation,yang2023radiative_limited}. For quantum applications, however, the relevant quantity is often the radiative rate and branching ratio of one selected Stark-to-Stark ZPL. Optical cavities can selectively enhance such a transition through the Purcell effect and thereby reduce the measured total lifetime to tens of microseconds or, in strongly enhanced structures, a few microseconds; values around 10~$\mu$s have been reported for cavity-coupled LiNbO$_3$\cite{yang2023}, with a comparable microsecond-to-tens-of-microseconds range demonstrated in CaWO$_4$ nanophotonic devices\cite{ourari2023indistinguishable,uysal2025spin_photon}.

\section*{Results}

\subsection{Erbium related defects in talc.}

One layer of talc contains two silicate sheets, which consist of covalently bonded SiO$_4$ tetrahedra, see Fig.~\ref{fig:geo}.  Between the two silicate layers, magnesium (Mg$^{2+}$) and hydroxide (OH$^{-}$) ions are located, which bind the layers with ionic bonds. These quasi-2D layers are then stacked together with van der Waals forces forming bulk talc (not shown).

In our study we consider four types of erbium related defects: erbium substitutional defect at the magnesium sites (Er$_{\text{Mg-1}}$ and Er$_{\text{Mg-2}}$), erbium substitutional defect and hydrogen vacancy complex (V$_{\text{H}}$Er$_{\text{Mg-x}}$), erbium substitutional defect and two hydrogen vacancy complex (2V$_{\text{H}}$Er$_{\text{Mg-x}}$), and erbium interstitial defect (Er$_i$).

The magnesium sites, Mg$_1$ and Mg$_2$, see Fig.~\ref{fig:geo},  can be substituted with transition metal impurities. This is supported by our previously work,~\cite{dolecsek2025native} where we showed that the formation energy of a magnesium vacancy in talc is significantly lower than that of a silicon vacancy. 

As erbium prefers cationic form (predominantly Er$^{3+}$) in nature, it is reasonable to assume that it also occupies the cationic position (i.e., the Mg$^{2+}$ site) when applied as a dopant in talc. Two chemically nonequivalent Mg positions can be distinguished in talc. The two Mg types differ in their local bonding environments, as visualized in Fig.~\ref{fig:geo}c. All Mg$^{2+}$ ions are coordinated in an octahedral fashion, by two oxygens from OH$^-$ ions and four oxygens from SiO$_4$ tetrahedrons. In talc strucutre there are two possibilities for the relative arrangement of the aforementioned six ligands. Mg$_1$ and Mg$_2$ represent the trans case (OH$^-$ ions in opposite position, at $\approx$ 180° of HO-Mg-OH angle) and the cis case (OH$^-$ ions in adjacent position, at $\approx$ 90° of HO-Mg-OH angle), respectively. Accordingly, Er$_{Mg1}$ and Er$_{Mg2}$ defects must be treated separately.

Although the hydroxyl groups of talc are not generally labile under ambient conditions, proton removal may become relevant under growth, irradiation, annealing, or charge-compensation conditions close to defects. This is consistent with the anisotropic surface chemistry of talc, where acid-base activity is mainly associated with edge sites rather than the basal planes, and with infrared studies showing proton migration and dehydroxylation upon heating.\cite{yan2011anisotropic,zhang2006dehydroxylation} We therefore consider Er$_{Mg}$ complexes with one or two adjacent hydrogen vacancies ($V_H$), in addition to the hydroxyl-intact configuration.

Altogether, we arrive at a total of six erbium defect configurations studied in different charge states, summarized in Table~\ref{tab:nomenclature}.   Different charge states result in different formal charge of the erbium cation, which fundamentally affects the optical properties (\emph{vide infra}). Therefore, we put a specific emphasis on erbium charge in our nomenclature; for example, V$_H$Er$^{3+}_{\text{Mg}2}$ indicates a neutral defect, as the absence of an Mg$^{2+}$ ion and a $H^+$ ion compensates the triple positive charge of Er$^{3+}$, see Table~\ref{tab:nomenclature}. Analogously, if the overall charge is positive or negative, one unit of charge is added (V$_H$Er$^{4+}_{\text{Mg}2}$) or removed (V$_H$Er$^{2+}_{\text{Mg}2}$) from Er, as the formal charge of main-group ions remains unaffected.

We further note that the overall spin quantum number (S) of the defect is solely determined by the charge state of the erbium cation, as indicated in Table~\ref{tab:nomenclature}.

Finally, in addition to the above 6 intralayer defects, we considered erbium impurities intercalated between two van der Waals-bound layers of talc ($Er_i$) as the 7th possible configuration.

\begin{table*}[t]
\begin{center}
 \caption{List of erbium defects and charge states studied in this work. The nomenclature emphasizes the formal charge of erbium as well as the spin of the defects.}
 \begin{tabular}{ |c | c| c| c | c | c|}
 \hline
Mg site & \# of $V_H$ & Defect configuration & \multicolumn{3}{c|}{Charge state}  \\[0.5ex]
 &&& q = 1  & q = 0  & q = -1  \\
 \hline
1 & 0 & Er$_{\text{Mg}1}$ & $\mathbf{Er^{3+}_{Mg1}}$ & $\mathbf{Er^{2+}_{Mg1}}$ & $\mathbf{Er^{+}_{Mg1}}$ \\
&  &  & S = 3/2 & S = 1 & S = 3/2\\
\hline
1 & 1 & V$_H$Er$_{\text{Mg}1}$ & $\mathbf{V_HEr^{4+}_{Mg1}}$ & $\mathbf{V_HEr^{3+}_{Mg1}}$ & $\mathbf{V_HEr^{2+}_{Mg1}}$ \\
&  &  & S = 2 & S = 3/2 & S = 1\\
\hline
1 & 2 & 2V$_{\text{H}}$Er$_{\textbf{Mg1}}$ & $\mathbf{2V_HEr^{5+}_{Mg1}}$ & $\mathbf{2V_HEr^{4+}_{Mg1}}$ & $\mathbf{2V_HEr^{3+}_{Mg1}}$ \\
&  &  & S = 5/2 & S = 2 & S = 3/2\\
\hline
2 & 0 & Er$_{\text{Mg2}}$ & $\mathbf{Er^{3+}_{Mg2}}$ & $\mathbf{Er^{2+}_{Mg2}}$ & $\mathbf{Er^{+}_{Mg2}}$ \\
&  &  & S = 3/2 & S = 1 & S = 3/2\\
\hline
2 & 1 & $V_HEr_{Mg2}$ & $\mathbf{V_HEr^{4+}_{Mg2}}$ & $\mathbf{V_HEr^{3+}_{Mg2}}$ & $\mathbf{V_HEr^{2+}_{Mg2}}$ \\
&  &  & S = 2 & S = 3/2 & S = 1\\
\hline
2 & 2 & $2V_HEr_{Mg2}$ & $\mathbf{2V_HEr^{5+}_{Mg2}}$ & $\mathbf{2V_HEr^{4+}_{Mg2}}$ & $\mathbf{2V_HEr^{3+}_{Mg2}}$ \\
&  &  & S = 5/2 & S = 2 & S = 3/2\\
\hline

\end{tabular}
\label{tab:nomenclature}
\end{center}
\end{table*}

\subsection{Defect strucutres.}

After the geometry relaxation of pristine talc, the Mg-O bond lengths in the octahedral coordination environments of the $\text{Mg}_1$ and $\text{Mg}_2$ sites are approximately equal in all directions, and are determined to be $2.04-2.07\text{\AA}$. The value of trans and cis $\text{O-Mg}_1$\text{-O} angles is $179^\circ-180^\circ$ and $84^\circ-95^\circ$, respectively. The deviation from the ideal octahedral angles $180^\circ$ and $90^\circ$ derive from the intrinsic asymmetry of the lattice.

\begin{figure*}
\includegraphics[width=\textwidth]{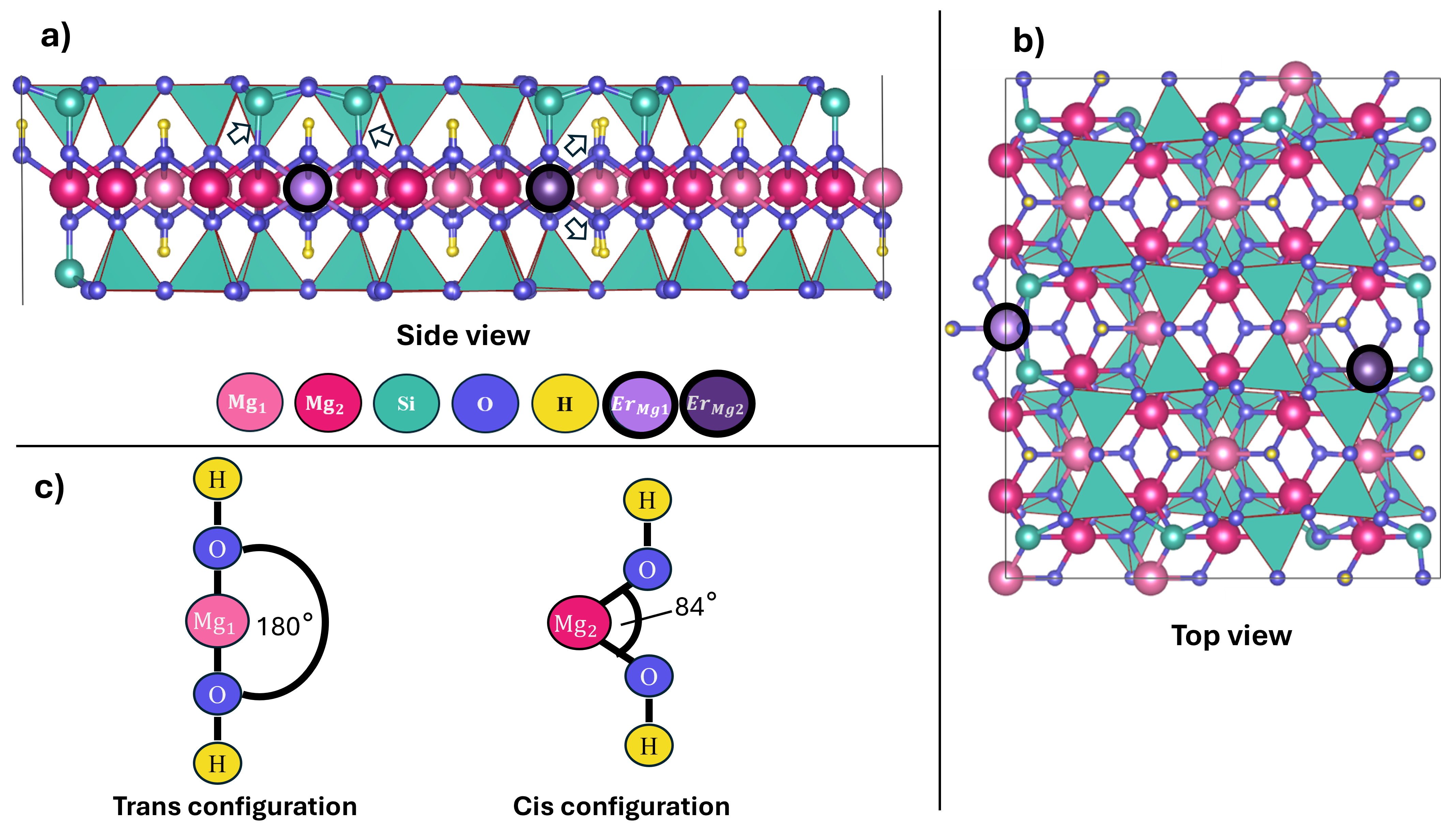}
	\caption{Structure of single-layer talc doped with Er atoms in a) side view, revealing two sheets of silicates (SiO$_4$) on the top and on the bottom of the structure, and b) in top view. For clarity, $SiO_4$ tetrahedra are filled with green color, while the position of individual Si atoms is only shown for selected sites. c) Visualization of the conceivable relative configurations of OH ligands in the coordination sphere of Mg$_1$ and Mg$_2$ atoms.}
	\label{fig:geo}  
\end{figure*}

Interestingly, the erbium-doped talc undergoes only minor structural modifications, see Fig.~\ref{fig:geo}. Following Er incorporation to either ${\text{Mg}_1}$ or ${\text{Mg}_2}$ site, the metal-oxygen bond lengths generally increase to approximately $2.2\text{\AA}$ - $2.3\text{\AA}$, albeit 1-2 out of the six Er-O bonds tends to slightly shrink to 2.01 - 2.04\AA. We note that the variation in Er-O interatomic distances can be partly traced back to the asymmetrically filled f orbitals. The metal-oxygen bond elongation is primarily driven by the larger ionic radius of erbium compared to magnesium. 

The O-Er-O bond angles also cover a significantly wider range than the O-Mg-O angles of pristine talc. In particular, trans O-Er-O angles of as low as $170^\circ$ are observed (the maximal value is still near $180^\circ$), while the cis angle is also versatile ($82^\circ-98^\circ$).   

A further detectable geometrical change induced by $Er_{Mg}$ substitution is the tilt of previously vertical Si-O bonds and OH groups, see white arrows in  Fig. \ref{fig:geo}. These distortions, however, possess negligible influence on the electronic properties of the lattice.

\subsection{Formation energy.}

Next, we study the formation energy ($E^f$) of the selected erbium related   erbium-related defects. The formation energy of defect $X$ in charge state $q$ can be written as a function of the Fermi level ($E_{\text{F}}$) as   \cite{freysoldt2014first}

\begin{equation}
    E^f[X^q] = E_{\text{tot}}[X^q] - E_{\text{tot}}[\text{bulk}]-\sum_i n_i \mu_i +qE_{\text{F}} + E_{\text{corr}} \text{,}
\label{eq:formation_energies}   
\end{equation}
where $E_{\text{tot}}[\text{bulk}]$ is the total energy of the pristine talc supercell, $E_{\text{tot}}[X^q]$ is the total energy of the defective talc supercell in charge $q$, while $n_i$ is the signed number of added or removed atoms, which possess a chemical potential of $\mu_i$ in their most stable elementary form. Finally, $E_{\text{corr}}$ stands for the charge correction term for $q \neq 0$ due to the periodic boundary condition.~\cite{freysoldt2018first}

First, we investigate the stability of the interstitial positively charged erbium impurities, intercalated between the talcum layers (Er$_{i}$; see Supplementary Information, section~S1.2). We found that such defects are unlikely to be detectably present due to their large formation energy. Even under the most favorable p-type conditions, the formation obtained with the R2SCAN exchange correlation functional is close to $10$~eV and rapidly increases with increasing Fermi energy. Therefore, we disregard Er$_{i}$ type defects and consider only Er$_{\text{Mg}}$ related defects hereinafter.

Next, we discuss the formation energies of neutral Er$_{\text{Mg}}$ related configurations, which are summarized in Table~\ref{tab:formation_energy}.  The R2SCAN formation energies are also compared with HSE06 results in Supplementary Note~S1.1, which show good agreement between the two methods. The obtained energies are remarkably low, reaching even negative values (from -0.5 to -3.3 eV), indicating that the incorporation of an erbium atom into a talc layer is energetically favourable. In addition, the calculated $E^f$ values show that the difference in formation energies at Mg$_1$ and Mg$_2$ sites is minimal. Consequently, Er$_{\text{Mg1}}$ and Er$_{\text{Mg}2}$ substitutions are expected to appear simultaneously in talc.  

\begin{table}[h!]
\begin{center}
 \caption{Neutral-charge formation energies ($E^f$) and charge transition levels ($\epsilon$, given relative to the valence band maximum) of defect structures considered in our study. All values are obtained with the METAGGA R2SCAN level of theory in a single-layer talcum model. (*R2SCAN optimization for the charge-neutral structure failed. The provided values are estimations based on HSE06 test calculations, see SI, section~S1.1)}
 \begin{tabular}{| c || c||c|c|c|}
 \hline
 Defects & $E^f$(q=0) & $\epsilon$(+2/+1) & $\epsilon$(+1/0)  & $\epsilon$(0/-1)\\[0.5ex]
   &  (eV) &  (eV) &  (eV) & (eV) \\[0.5ex]
 \hline
$Er$$_{\text{Mg1}}$  & -1.6* & -0.2 & 5.8*  & - \\
$Er$$_{\text{Mg2}}$   & -1.6 & -0.2 & 5.8 & - \\
\hline
$V_{\text{H}}$$Er$$_{\text{Mg1}}$   &  -3.3 & - &  0.9 & - \\
$V_{\text{H}}$$Er$$_{\text{Mg2}}$   &  -3.3 & - & 0.9  & - \\
\hline 
$2V_{\text{H}}$$Er$$_{\text{Mg1}}$ &  -0.9 & - & - &  2.5 \\ 
$2V_{\text{H}}$$Er$$_{\text{Mg2}}$ & -0.5 &  &  0.4  &  2.0 \\
\hline
\end{tabular}
\label{tab:formation_energy}
\end{center}
\end{table}

\begin{figure*}
\includegraphics[width=\textwidth]{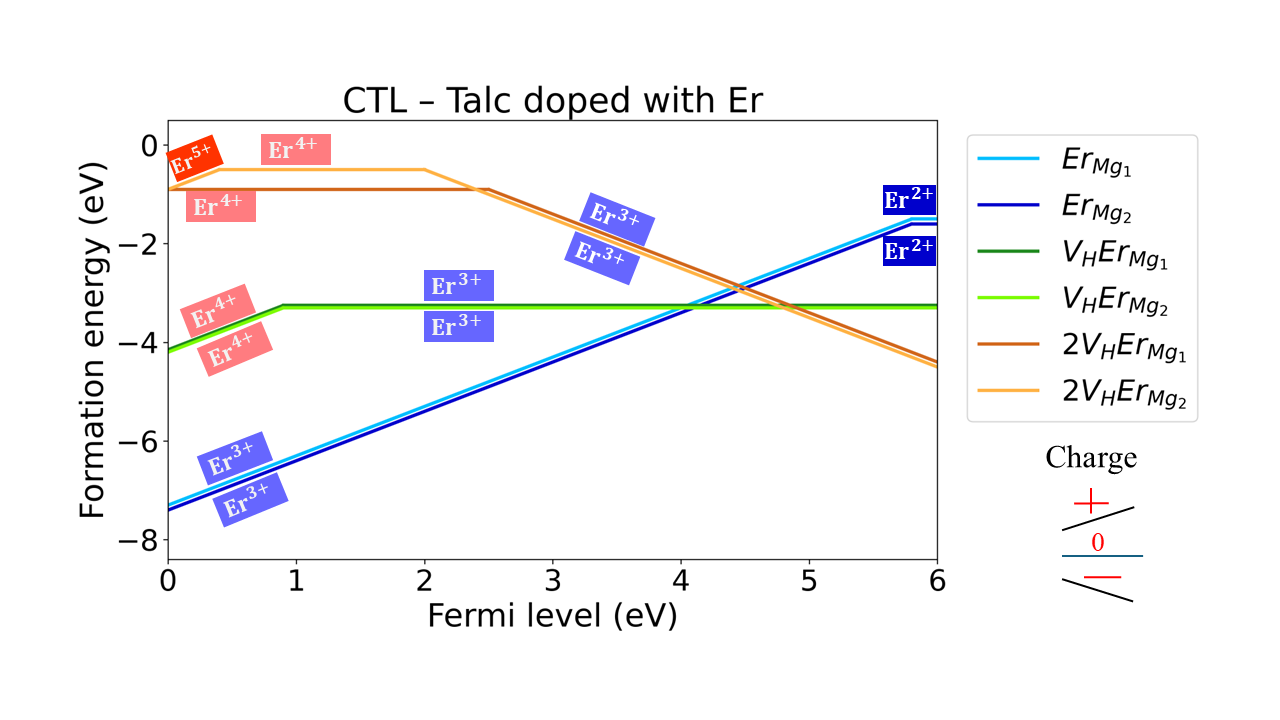}
	\caption{Formation energy of erbium defects as a function of the Fermi level ($E_F = 0$ corresponds to the valence-band maxiumum.)}
	\label{fig:formation_energy}  
\end{figure*}

As for non-neural charge states, the $qE_F$ term in Eq.~\ref{eq:formation_energies} varies in a wide interval, owing to the large band gap of talc\cite{dolecsek2025native}. Therefore, the formation energies can be further lowered under favorable circumstances, even by several electronvolts. This effect is visualized in Fig.~\ref{fig:formation_energy}, where $E^f$ is depicted with respect to the Fermi level. At each $E_F$, only the minimal-energy charge state is shown. 

Charge transition levels are calculated for all the considered defects using the formula\cite{weston2018native} of
\begin{equation}
    \epsilon(q|q^\prime)= \\ \frac{E_{\text{tot}}[X^q] + E_{\text{corr}}(q) - (E_{\text{tot}}[X^{q^\prime}]+E_{\text{corr}}(q^\prime))}{q^\prime-q}-\epsilon_{VBM}
\end{equation}
where $E^f(X^q)$ is the formation energy of defect $X$ in charge state $q$.

The computed values, including charge corrections, are provided in Table~\ref{tab:formation_energy}.
Analyzing the position of the charge transition level with respect to the position of the band edge, the valence band maximum in our case, we can identify the stable charge states of the defects at a given value of the Fermi energy, see Fig.~\ref{fig:formation_energy}.

Conspicuously, while V$_{\text{H}}$Er$_{\text{Mg}}$ defects favor electroneutrality regardless of the Fermi level (except for extreme p-type conditions), the most stable charge state of Er$_{\text{Mg}}$ and 2V$_{\text{H}}$Er$_{\text{Mg}}$ is positive and negative, respectively. In the case of Er$_{\text{Mg}}$, the (+1/0) charge transition appears at n-type limit at 5.8~eV above the valence band edge, while $E^f$ lowers to  $\approx-7$ eV at p-type talc when the Fermi energy is close to the VBM. Furthermore, we find that the +2 charge state of Er$_{\text{Mg}}$ is not stable in the band gap of talc, see Table~\ref{tab:formation_energy}. For 2V$_{\text{H}}$Er$_{\text{Mg}}$, however, the (0/-1) charge transition occurs at intermediate $E_F$ levels, making the neutral $q = 0$ state also relevant at weak p-type conditions. For semi-insulating and n-type samples, the negative charge state is stable. We also note that  CTLs slightly depend on the position of Er in the lattice in the case of 2V$_{\text{H}}$Er$_{\text{Mg}}$ configuration, see Fig.~\ref{fig:formation_energy}. 

Based on Table~\ref{tab:nomenclature} and Fig.~\ref{fig:formation_energy}, we can relate the most stable charge states to the formal charge of the erbium cation. Chemical intuition suggests the commonly observed Er$^{3+}$ to be dominantly formed, which is corroborated by the above results. The positively charged Er$_{\text{Mg}}$, the neutral V$_{\text{H}}$Er$_{\text{Mg}}$ and the negatively charged 2V$_{\text{H}}$Er$_{\text{Mg}}$ defects all correspond to Er$^{3+}$, see Fig.~\ref{fig:formation_energy}. The Er$^{3+}_{\text{Mg}}$ V$_{\text{H}}$Er$^{3+}_{\text{Mg}}$ and 2V$_{\text{H}}$Er$^{3+}_{\text{Mg}}$  configurations are the stable ones for a Fermi energy positioned in the middle of the band gaps expected for pristine talc layers lacking intrinsic defects\cite{dolecsek2025native}. Notable alternative cationic forms of Er were found for neutral 2V$_{\text{H}}$Er$_{\text{Mg}}$ (corresponding to Er$^{4+}$ ion at the center), and - at the extreme limits of $E_F$ - for neutral Er$_{\text{Mg}}$ (Er$^{2+}$ center), for positively charged V$_{\text{H}}$Er$_{\text{Mg}}$ (Er$^{4+}$ center) and for positively charged 2V$_{\text{H}}$Er$_{\text{Mg}}$ (Er$^{5+}$ center).

\subsection{Kohn-Sham electronic structure of the most stable defects.}

Next, we discuss the electronic structure and the optical properties of the erbium defect forms found to be stable at Fermi levels lying within the band gap  (Er$^{2+}_{\text{Mg}}$, Er$^{3+}_{\text{Mg}}$, V$_{\text{H}}$Er$^{3+}_{\text{Mg}}$, V$_{\text{H}}$Er$^{4+}_{\text{Mg}}$,2V$_{\text{H}}$Er$^{3+}_{\text{Mg}}$, 2V$_{\text{H}}$Er$^{4+}_{\text{Mg}}$, and  2V$_{\text{H}}$Er$^{5+}_{\text{Mg}}$), at both possible Mg positions.

\begin{figure*}
\includegraphics[width=\textwidth]{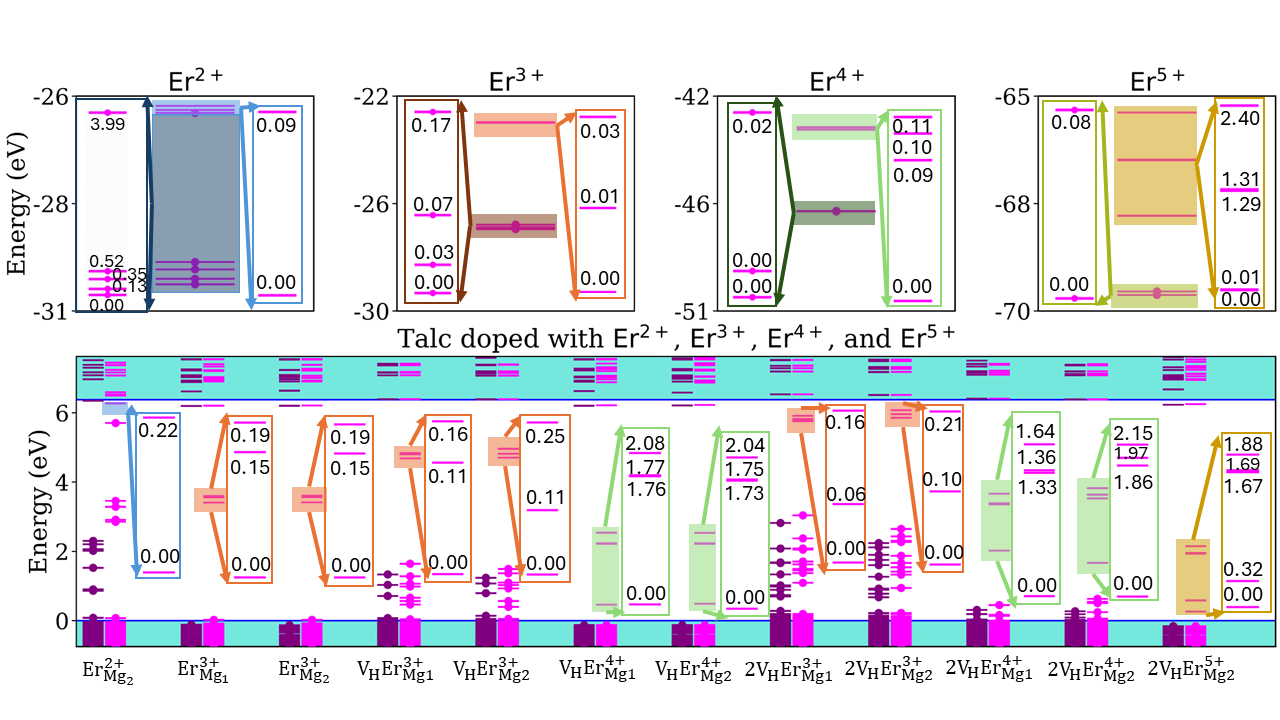}
	\caption {Kohn-Sham electronic structure of the thermodynamically most stable erbium impurities in talc, compared to the 4f subshell of free erbium cations (upper row of diagrams). Purple and pink lines represent spin-up and  spin-down channells, respectively, while circles indicate occupation by an electron. Bottom and top green regions indicate the valence and conduction band of talc, respectively.}
	\label{fig:Kohn-Sham}  
\end{figure*}

The single-determinant (Kohn-Sham) electronic structure of the most stable erbium-containing defects is shown in Fig.~\ref{fig:Kohn-Sham}. For the case of Er$^{3+}_{\text{Mg}}$, the position of conduction and valence band orbitals closely resembles the Kohn-Sham structure of the bulk talc (see green bands), while for V$_{\text{H}}$ containing cases the valence band edge can be shifted upward by 1-2 eV due to the perturbation of the narrow, large density of states oxygen related valnece band, which shows high susceptibility for the pertubation of oxigen occupany\cite{dolecsek2025native}.

In the band gap, only three (four) empty spin-down orbitals can be found for Er$^{3+}$ (Er$^{4+}$),  which correspond to the unoccupied $4f$ atomic orbitals of Er affected by a small degree of hybridization. The remaining $4f$ orbitals are situated deep inside the valence band.  

It is interesting to compare the arrangement of the aforementioned empty $4f$ orbitals to the structure of free erbium cations in vacuum.
%, which is visualized in Fig.~\ref{fig:Kohn-Sham_zoom}
If Mg$_1$ or Mg$_2$ is substituted by Er$^{3+}$ in monolayer talc (Er$^{3+}_{\text{Mg}}$, V$_{\text{H}}$Er$^{3+}_{\text{Mg}}$, and 2V$_{\text{H}}$Er$^{3+}_{\text{Mg}}$), the cristal field induces orbital splitting only $6$-$8$ times larger than that of the free Er$^{3+}$ ion. The splitting of the empty $4f$ orbitals reaches 0.15-0.25 eV, depending on the configuration. This result for Er$^{3+}_{\text{Mg}}$, V$_{\text{H}}$Er$^{3+}_{\text{Mg}}$ and 2V$_{\text{H}}$Er$^{3+}_{\text{Mg}}$ implies that such systems can be interpreted as practically free Er$^{3+}$, i.e.\ no $4f$ related bonds are formed. By contrast, the $4f$ orbitals of Er$^{4+}$ defects (V$_{\text{H}}$Er$^{4+}_{\text{Mg}}$ and 2V$_{\text{H}}$Er$^{4+}_{\text{Mg}}$) cover an energy range of 1.6-2.2 eV, which are 15 - 20 times larger than that of the splitting of the free Er$^{4+}$ ion, indicating significant interaction between the ion and the surrounding ligands. This enhanced susceptibility to the ligand field, which can presumably be traced back to the higher effective nuclear charge of Er$^{4+}$ acting on the neighboring oxygen atoms, is also manifested in the Mg site dependence of orbital energy levels. In particular, for 2V$_{\text{H}}$Er$^{4+}_{\text{Mg1}}$, a significantly smaller energy separation is observed (1.64 eV) than for 2V$_{\text{H}}$Er$^{4+}_{\text{Mg2}}$ (2.15 eV). 

\subsection{Optical properties of Er$^{3+}_{\text{Mg}}$ and Er$^{4+}_{\text{Mg}}$ defects.}

The $^4I_{13/2} \rightarrow ^4I_{15/2}$ optical transition of Er$^{3+}$ can be modeled by WFT, in the framework of quasi-degenerate perturbation theory (QDPT)~\cite{Roemelt_2013}. At first, we generate 13 spin-free eigenstates (the $I$ states are 2L+1=13-fold degenerate) in a complete active space - self-consistent field (CASSCF) calculation~\cite{Olsen2011_CASSCFPerspective}, see Methods and Supplementary Note~S1.3 for details. By introducing spin-orbit coupling (SOC), these 13 states are split into 13$\times$4=52 coupled basis states. The linear combination of this basis states gives the proper wavefunction of the individual, (2J+1)-fold degenerate spin-orbit coupled eigenstates of $^4I_{J}$. We note that the transition for the Er$^{4+}$ ion, which has 10 electrons and 4 holes on its f orbitals corresponding to $S=2$ and $L=6$, is ${^5}I_{7} \rightarrow {^5}I_{8}$.

In free Er$^{3+}$ and Er$^{4+}$ ions, the (2J+1)-fold degeneracy of spin-orbit coupled sublevels is exact, and the optical transition is only allowed by a weak magnetic dipole moment ($\mu_M$). In this case, the radiative rate ($k_{PL,M}$) can be calculated as
\begin{equation}
    k_{\mathrm{PL,M}} = \frac{1}{\tau_{PL, M}}  = \sum_{if}p_i\frac{n^3 \Delta E_{if}^3}{3 \pi \varepsilon_0 \hbar^4 c^5} \left| \mathbf{\mu}_{M,if} \right|^2 \text{,}
\label{eq:k_PL_M}
\end{equation}
where $\Delta E_{if}$ is the energy gap between the initial (i) and final (f) spin-orbit coupled eigenstate and $p_i$ is the Boltzmann factor of the initial state. We recall that the number of possible ${if}$ combinations is ($2J_i+1$)($2J_f+1$). 

To obtain reference values for studying the erbium ions in talc, we first modeled the emission rate for Er$^{3+}$ and Er$^{4+}$ in vacuum (refractive index: $n$ = 1 ). The results are shown in the corresponding rows of Table~\ref{tab:PL}, where we gathered the energy difference related to J quantum number change ($\Delta E$), a Boltzmann-averaged representative magnetic dipole ($\mu_M$, see also in Supplementary Information section~1.5 for the details of Boltzmann statistics) and the lifetime of emission ($\tau_{PL}$).   

It can be observed that a photon energy of 0.75 eV was obtained for free E$r^{3+}$ using our WFT protocol, which is in good agreement with the experimental observation of 1550 nm (0.80 eV). Moreover, the radiative lifetime of 121 ms is consistent with $4f^n$-specific intermediate coupling calculations~\cite{Dodson2012}, which gave 98 ms for the same transition. As for Er$^{4+}$, the transition was found to occur with comparable characteristics, albeit with a slightly lower photon energy (0.63 eV, or 1970 nm) and also a lower rate (162 ms lifetime). 

In the next step, we discuss the effect of the crystal field of the embedding talc lattice on these characteristic emissions. Firstly, due to the presence of the external crystal field, states of equal $J$ quantum numbers lose their degeneracy and are split into Stark components. Consequently, multiple peaks are expected in the photoluminescence spectrum, lying in an energy range of 30-120 meV of width (considering an intensity cutoff of 95\%).
%, rather than at a well defined wavelength.
Furthermore, the weighting by $p_i$ in Eq.~\ref{eq:k_PL_M} cannot be substituted by a simple arithmetic averaging anymore, and the results become sensitive to temperature. In this work, we characterized the emission at room temperature (300 K). 

Secondly, $k_{\text{PL,M}}$ scales with the third power of the refractive index, therefore, the talc environment with $n$=1.55 significantly accelerates the emission. $k_{\text{PL,M}}$ can also be altered due to an environment-induced change in the $\mu_{M,if}$ terms, however, the sign of this effect cannot be predicted \emph{a priori}.

Thirdly, the electric dipole moment ($\mu_E$) can also arise in the lattice, which contributes to the PL rate as
\begin{equation}
    k_{\mathrm{PL,E}} = \frac{1}{\tau_{PL,E}} = \sum_{if}p_i\frac{n \Delta E_{if}^3}{3 \pi \varepsilon_0 \hbar^4 c^3} \left| \mathbf{\mu}_{E,if} \right|^2
\end{equation}
producing an overall half-life of 
\begin{equation}
    \tau_{\mathrm{PL}} = \frac{1}{k_{PL,M}+k_{PL,E}} \text{.}
\end{equation}

The results obtained for Er$^{3+}$ and Er$^{4+}$ ions, substituted to the Mg$_1$ or Mg$_2$ site of talc, are also provided in Table~\ref{tab:PL}. Apart from the characteristics discussed for free ions, we also present here the Boltzmann-averaged representative transition dipole moment ($\mu_E$) and the contribution of the electric and magnetic terms to the overall rate in percentage (M\% and E\%). 

Since we have already observed the vast robustness of the Kohn-Sham structure of Er$^{3+}$ against crystal field effects, it is not surprising to see that all talc-embedded Er$^{3+}$ systems share practically the same optical lifetime (27-31 ms) for $^4I_{13/2} \rightarrow ^4I_{15/2}$ relaxation. The enhanced rate of photoemission relative to the parent ion ($\tau_{\text{PL}}$=121~ms) can almost solely be traced back to the introduction of the lattice refractive index (n = 1.55) to the $k_{\text{PL,M}}$ term, which gives a factor of $1.55^3 \approx 3.7$. Compared to this effect, the crystal field-induced change in the magnetic transition dipole moment (from 1.17 to 1.21-1.24 $\mu_B$) can be considered as insignificant. As a secondary effect, the appearance of a small ($<$ 0.01 Debye) electric transition dipole moment also contributes to the reduction of the radiative lifetime. For V$_{\text{H}}$Er$^{3+}_{\text{Mg}}$ and 2V$_{\text{H}}$Er$^{3+}_{\text{Mg}}$ defects, we observe approximately an order of magnitude increase of the electric transition dipole moment. As a consequence, the contribution of the electric dipole-driven transition rate increases to 11\%-15\%, which, however, showcases that the magnetic dipole transition still dominates. An exception to this trend is the 2V$_{\text{H}}$Er$^{3+}_{\text{Mg1}}$ configuration, where, due to the trans configuration of the hydrogen vacancies, see Fig.~\ref{fig:geo}, the V$_{\text{H}}$ induced electric field perturbations cancel out to a large degree, see Table~\ref{tab:PL}. 

In the case of the $^5I_{7} \rightarrow ^5I_{8}$ emission in Er$^{4+}$, the crystal field effects are more prevalent, as expected from f-orbital splittings seen in the Kohn-Sham orbital picture in Fig.~\ref{fig:Kohn-Sham}. We observe in Table~\ref{tab:PL} that the talc environment considerably enhances the magnetic transition dipole moment. While $\mu_M$=1.30 $\mu_B$ was calculated for the free ion, 2V$_{\text{H}}$Er$^{4+}_{\text{Mg1}}$ shows a 1.5-fold enhancement ($\mu_M$=1.9 $\mu_B$). The Stark splitting is also more significant, as a result of which a broader range of ZPL energies appear (ZPLs of as low as 0.56 eV and as high as 0.68 eV appear, compared to the free-ion ZPL of 0.63 eV), the higher end of which further reduces the lifetime. These effects, combined with the refractive index-related acceleration factor ($n^3\approx3.7$) result in lifetimes of 29-36 ms. We note that the similarity to Er$^{3+}$ lifetimes, despite the aforementioned emission-facilitating crystal field effects, can be traced back to the longer-living parent ion (Er$^{4+}$, with $\tau_{\text{PL}} = 162$~ms).  

Altogether, the primary effect of the talc lattice is the facilitation of the magnetic dipole induced photoluminescence ($k_{\text{PL,M}}$) of the embedded Er$^{3+}$ and Er$^{4+}$ ions via its refractive index. In the case of Er$^{3+}$ and Er$^{4+}$, additional non-negligible effects arise from the enhancement of the electric transition dipole and the enhancement of the magnetic transition dipole, respectively, further decreasing the a radiative lifetime compared to the free ion. The obtained lifetimes fall in the range of 29-36 ms, which is commensurable to those measured experimentally in Er$^{3+}$-doped crystals reported in the literature~\cite{carnall1989systematic,miniscalco1991erbium,bottger2006spectroscopy,cornacchia2007growth,becker2025spectroscopic,fleuster1994optical,wang2020incorporation,yang2023radiative_limited}. Therefore, based on our results, the natural 2D-structured lattice of talc can be considered as a promising host for quantum technological applications of Er$^{3+}$ and Er$^{4+}$ photoemission lines.

To evaluate the performance of talc-embedded Er centers regarding quantum applications, where a single ZPL, i.e., one selected Stark transition, is targeted, we gathered the brightest expected single-photon emission channels in Table~\ref{tab:PL_Q}. We also assigned the magnetic sublevel ($M_J$ quantum number) to the initial and final Stark states involved in the selected optically favorable transition, even though we emphasize that presence of the crystal field inevitably mixes the free-ion-like $M_J$ sublevels and the unequivocal identification is therefore not always possible.
%We emphasize here that the lifetimes ($\tau_{\mathrm{PL}}$) in this table are weighted by the Boltzmann factor of the initial Stark state ($p_i$).
As expected, the maximal accessible single-photon rates - without Boltzmann weighting of the initial state- lie in the order of magnitude of the total PL rate (40-80 ms, see SI for details). Such slow optical decay is obviously inappropriate for practical single-photon-source operation and, therefore, any selected ZPL would have to be enhanced by optical engineering, most naturally by exploiting the Purcell effect. In this regard, however, talc has advantageous features: as an exfoliable van der Waals silicate, it should be compatible with heterogeneous integration onto pre-fabricated dielectric or plasmonic nanophotonic structures, allowing the Er$^{3+}$ center to be placed in the near field of an optical resonator without aggressive nanofabrication of the host crystal itself. Analogous strategies have already been demonstrated for hBN single-photon emitters coupled to photonic-crystal and microring cavities and, more directly for rare-earth systems~\cite{Aharonovich2016,Kim2018,Froch2020,parto2022,dibos2018,Merkel2020,yang2023}.

\begin{table*}[t]
\begin{center}
 \caption{Optical properties of talc-embedded Er centers from the point of view of \textbf{general telecom-band optoelectric applications} (i.e. the total radiative decay involving all Stark transitions is investigated). Energy difference range ($\Delta E$), representative magnetic transition dipole moment ($\mu_M$), representative electric transition dipole moment ($\mu_E$), radiative lifetime ($\tau_{PL}$) and the contribution ratio of magnetic-iduced ($M\%$) and electric-induced ($E\%$) emission, calculated for the $^4I_{13/2} \rightarrow ^4I_{15/2}$ transition of Er$^{3+}$ and the $^4I_{7} \rightarrow ^4I_{8}$ transition of Er$^{4+}$ at WFT level.}
 \begin{tabular}{| c | c| c| c ||c||c|c|}
 \hline
 System & $\Delta E$ [eV] &$\mu_M$ [$\mu_B$] & $\mu_E$ [D] & $\tau_{PL}$ [ms] & M\% & E\% \\[0.5ex]
\hline
 Free Er$^{3+}$ &0.75&1.17&-&121&100\%&-\\
 \hline
 Talc, Er$^{3+}$$_{\text{Mg1}}$ &0.73-0.76&1.22&0.00046&31&99.9\%&0.1\% \\
 Talc, Er$^{3+}$$_{\text{Mg2}}$ &0.73-0.76&1.22&0.00098&31&99.7\%&0.3\% \\
 Talc, $V_HEr^{3+}$$_{\text{Mg1}}$ &0.73-0.78&1.21&0.00725&27&85.1\%&14.9\%\\
  Talc, $V_HEr^{3+}$$_{\text{Mg2}}$ &0.73-0.78&1.21&0.00663&28&87.4\%&12.6\%\\
 Talc, $2V_HEr^{3+}$$_{\text{Mg1}}$ &0.73-0.77&1.24&0.00024&31&100.0\%&0.0\% \\
 Talc, $2V_HEr^{3+}$$_{\text{Mg2}}$ &0.73-0.78&1.21&0.00600&29&89.3\%&10.7\% \\
 \hline
 Free Er$^{4+}$ &0.63&1.30&-&162&100\%&-\\
 \hline
 Talc, $2V_HEr^{4+}$$_{\text{Mg1}}$ &0.56-0.68&1.94&0.00044&29&100.0\%&0.0\% \\
 Talc, $2V_HEr^{4+}$$_{\text{Mg2}}$ &0.58-0.66&1.45&0.00629&36&91.7\%&8.3\%\\
 \hline

\end{tabular}
\label{tab:PL}
\end{center}
\end{table*}

\begin{table*}[t]
\begin{center}
 \caption{Optical properties of talc-embedded Er centers from the point of view of \textbf{quantum applications}, i.e., the highest radiative decay rate among all Stark transitions is investigated). Energy difference ($\Delta E$), initial ($M_{J_i}$) and final ($M_{J_f}$) J sublevels, magnetic transition dipole moment ($\mu_M$), and electric transition dipole moment ($\mu_E$). %Boltzmann distribution corrected radiative lifetime ($\tau_{PL}$) and the contribution ratio of magnetic-iduced ($M\%$) and electric-induced ($E\%$) emission is shown for the selected Stark transition. (We note that in the case of Er$^{3+}$, Kramers pairs rather than indvidual SOC states are considered, see SI, section~\ref{SI-subsec:Boltzmann} for details
 }
 \begin{tabular}{| c | c| c| c| c| c ||c||c|c}
 \hline
 System & $\Delta E$ [eV] & $M_{J_i}$ & $M_{J_f}$ &$\mu_M$ [$\mu_B$] & $\mu_E$ [D] \\[0.5ex]
 % & $\tau_{PL}$ [ms] & M\% & E\% \\[0.5ex]
\hline
 Talc, Er$^{3+}$$_{\text{Mg1}}$ &0.75& $\pm 13/2$& $\pm 13/2$&0.91&0.00016 \\
 %&224&100.0\%&0.0\% \\
 Talc, Er$^{3+}$$_{\text{Mg2}}$ &0.75 & $\pm 13/2$& $\pm 13/2$ &0.93&0.00018 \\
 %&215&100.0\%&0.0\% \\
 Talc, $V_HEr^{3+}$$_{\text{Mg1}}$ &0.73 & $\pm 13/2$& $\pm 15/2$ &0.98&0.00342\\
 %&196&94.9\%&5.1\%\\
  Talc, $V_HEr^{3+}$$_{\text{Mg2}}$ &0.73 & $\pm 13/2$& $\pm 15/2$ &0.99&0.00315 \\
  %&239&93.5\%&6.5\% \\
 Talc, $2V_HEr^{3+}$$_{\text{Mg1}}$ &0.76 & $\pm 5/2$& $\pm 7/2$ &1.04&0.00007 \\
 %&130&100.0\%&0.0\% \\
 Talc, $2V_HEr^{3+}$$_{\text{Mg2}}$ &0.75 & $\pm 13/2$& $\pm 15/2$ &1.05&0.00244 \\
 %&144&97.5\%&2.5\% \\
 \hline
 Talc, $2V_HEr^{4+}$$_{\text{Mg1}}$ &0.68 & $\pm 7$& $\pm 8$ &1.05&0.00009 \\
 %&136&100.0\%&0.0\% \\
 Talc, $2V_HEr^{4+}$$_{\text{Mg2}}$ &0.66 & $\pm 3$& $\pm 4$& 0.87&0.00366 \\
 %&466&96.5\%&3.5\% \\
 \hline

\end{tabular}
\label{tab:PL_Q}
\end{center}
\end{table*}

\section*{Discussion}

%We have investigated the properties of talc as a potential host for the quantum applications of the characteristic photoemission lines of erbium cations. Our quantum chemical calculations reveal that the substitution of $Mg_1$ or $Mg_2$ sites by Er occurs not only with low formation energies ($< 2 eV$ / site) but is often even thermodynamically favorable ($E_f$ is in the range of (-3)-(-5) eV for the most stable defects at intermediate Fermi levels.) The charge state of the formed defects, taking the possibility of nearby $V_H$ sites into account, corresponds to a formal Er$^{3+}$ or Er$^{4+}$ ion at the center. In both cases, telecom- or near-telecom-wavelength photoluminescence is expected from the talc-embedded erbium, wich occurs with 29-36 ms excited-state lifetime (summed for all Stark components). Considering the PL rate commensurable to that in commercially available Er-based optical devices and the advantages of the high-band-gap, natural 2D lattice of talc, we suggest that erbium-doped talc should be considered as a promising platform for telecom-band optoelectronic technologies. For quantum applications, the utilization of the single-photon emission arising from individual Stark transitions requires Purcell engineering, as the lifetime of individual optical decays falls in the range of hundreds of milliseconds. 

The present results establish talc as a previously unexplored rare-earth host that combines favorable defect thermodynamics with optical characteristics relevant for telecom photonics and quantum technologies. In contrast to many conventional Er-doped materials, talc provides a naturally layered van der Waals lattice that can be mechanically exfoliated down to single layer level and integrated into hybrid photonic architectures without the need for epitaxial growth. The calculated formation energies indicate that substitutional incorporation of erbium is energetically favorable and should occur readily under suitable synthesis or diffusion conditions. Importantly, the preferred charge states correspond predominantly to Er$^{3+}$, ensuring access to the technologically important ${^4}I_{13/2} \rightarrow {^4}I_{15/2}$ transition near 1.55~$\mu$m.

A key observation is that the talc crystal field perturbs the Er 4f shell only weakly in the Er$^{3+}$ configuration. Consequently, the telecom transition largely retains its free-ion character, while the local symmetry generates Stark splitting that provides multiple spectrally resolvable optical transitions. This combination is particularly attractive for quantum photonics because it enables narrow optical lines while preserving the exceptional robustness of rare-earth 4f states. The calculated radiative lifetimes of approximately $\sim$30~ms fall within the range reported for established rare-earth photonic hosts, suggesting that the talc environment does not compromise the desirable optical properties of erbium.

From a device perspective, the most significant advantage of talc may arise from its compatibility with heterogeneous integration. Exfoliated talc flakes containing Er centers could be transferred onto pre-fabricated photonic-crystal cavities, microring resonators, dielectric metasurfaces, or plasmonic nanostructures. Such an approach avoids aggressive nanofabrication of the host crystal itself while still enabling large Purcell factors and efficient control of selected Stark transitions. Similar integration strategies have proven highly successful for other van der Waals quantum emitters and could provide a realistic route toward telecom single-photon sources based on erbium centers in talc.

The natural two-dimensional character of talc also opens opportunities beyond isolated emitters. Layered erbium-doped talc could be incorporated into van der Waals heterostructures together with graphene, hexagonal boron nitride, transition-metal dichalcogenides, or lithium-niobate-on-insulator photonic circuits. Such hybrid platforms may enable optical modulation, electrical tuning, and cavity-enhanced readout of rare-earth states while preserving the scalability associated with layered materials. Because talc possesses atomically flat, chemically inert surfaces and a wide band gap, it is expected to introduce minimal optical loss and limited background luminescence.

Regarding fabrication, several experimentally accessible routes seem feasible. Er incorporation may be achieved during growth, through ion implantation followed by thermal annealing, or via diffusion-assisted doping protocols already established for rare-earth ions in oxides and silicates. The large thermodynamic driving force predicted here suggests that stable substitutional configurations should form readily once Er reaches Mg lattice sites. Future experimental work should clarify achievable concentrations, optical linewidths, spectral diffusion characteristics, and spin coherence properties.

Overall, our results suggest a broader perspective for naturally occurring layered silicates that represent a largely unexplored family of quantum-photonic materials. Within this emerging class, erbium-doped talc appears as a particularly promising candidate that combines telecom-band emission, defect stability, and van der Waals integrability. These features position it as a potentially attractive platform for future integrated quantum-photonic technologies.

\section*{Methods}

In this work, both density functional theory and wavefunction theory were used to study the properties of Er-doped talc. 

\subsection{Density functional theory on supercells.}

To investigate formation energies and Kohn-Sham electronic structures, we employ density functional theory calculations in periodic boundary conditions as implemented in the Vienna Ab initio Software Package (VASP, version is VASP/5.4.4)\cite{VASP,VASP2}. For the present work, the R2SCAN\cite{Furness2020_r2SCAN} density functional is selected, which is computationally less demanding than the widely applied hybrid functional of HSE06\cite{Heyd2003_HSE,Krukau2006_HSE06,Heyd2006_HSE_erratum}. Our test calculations, detailed in section~S1.1 of the Supplementary Information, revealed that R2SCAN sufficiently reproduces the HSE06-level geometries and formation energies of Er centers. To account for the weak van der Waals interaction between the layers, the Grimme-D3 correction\cite{grimme_consistent_2010} is applied on top of R2SCAN.  Unless indicated otherwise, single-layer talc supercells of $3 \times 2$  size (containing 252 atoms) are used, with only Gamma-point sampling of the Brillouin zone. 

Kohn-Sham wavefunctions of the valence electrons are expanded in a plane wave basis set of 440~eV. We use the projector-augmented wave (PAW) method\cite{PAW} to describe the potential of the nucleus and the weakly interacting core electrons of the atoms. A vacuum spacing of $9~\text{\AA}$ is introduced into the simulation cell to minimize van der Waals interactions between neighboring periodic supercells. Furthermore, to accurately describe the 4f shell, we turn on non-spherical exchange-correlation contributions inside the PAW augmentation spheres. The stopping criteria for the electronic self-consistent field (SCF) loop and the structural relaxation are $\Delta E < 10^{-6}$~eV and $F_i < 10^{-3}$~eV/\AA, where $\Delta E$ is the change of the total energy in an iteration of the SCF loop, and $F_i$ is the length of the force vector acting on atom $i$.

\subsection{Wavefunction theory on cluster models.}

The description of initial and final spin-obit coupled states of Er$^{3+}$ and Er$^{4+}$ photoluminescence lines requires multideterminantal wavefunction theory, which is, however, only available for cluster models in conventional quantum chemical software. Herein, based on the work of Tunega and Turi Nagy~\cite{Tunega1993TalcElectronStructure}, we construct hydrogen terminated clusters containing $298-n_{V_H}$ atoms, where $n_{V_H}$ is the number of hydrogen vacancy sites. A more detailed description and the visualization of these models is provided in the Supplementary Information under section~1.3. The position of atoms (with the exception of the terminating hydrogens) were taken from DFT-optimized supercell geometries.

To confirm that the finite size of the cluster does not introduce any artifacts, we compared the local electrosctatic potential in the vicinity of the Er dopant in the cluster and in the supercell at the same level of theory and found no significant differences (see  section~S1.4 of the Supplementary Information).  

Wave-function-theory (WFT) calculations were carried out using the ORCA 6.1.0 program package\cite{Neese2012_ORCA,Neese2020_ORCA,Neese2025_ORCA6}. Static correlation within the open 4f shell was treated at the state-averaged complete active space self-consistent field (SA-CASSCF) level\cite{Roos1980_CASSCF}. The active space comprised seven active orbitals, corresponding to the 4f subshell of Er, which were selected from initial DFT orbitals, transformed by the atomic valence active space (AVAS) method\cite{Sayfutyarova2017_AVAS}. The number of active electrons was chosen according to the formal oxidation state of the erbium ion: CAS(11,7) and CAS(10,7) were used for Er$^{3+}$ ($4f^{11}$ configuration) and Er$^{4+}$ ($4f^{10}$ configuration), respectively. In both cases, 13 spin-free CASSCF roots were requested, corresponding to the $(2L+1)=13$ components of the $I$ term ($L=6$). The spin multiplicity was set to quartet for Er$^{3+}$ and quintet for Er$^{4+}$, consistent with the $^4I$ and $^5I$ ground terms, respectively.

On the basis of the obtained spin-free CASSCF eigenstates, spin--orbit coupled states were calculated within the quasi-degenerate perturbation theory (QDPT) framework for multireference spin-Hamiltonian calculations\cite{Roemelt_2013,GanyushinNeese2006_ZFS,GanyushinNeese2013_CASSOC,Lang2025_SOCSSC}. In this approach, the spin--orbit coupling (SOC) and spin--spin coupling (SSC) operators are represented in the basis of the non-relativistic CASSCF roots ${\Psi_I}$ and their spin sublevels. The effective QDPT matrix can be written as
\begin{multline} \mel{\Psi_I^M}{\hat{H}_{CASSCF}+\hat{H}_{SOC}+\hat{H}_{SSC}}{\Psi_J^{M'}} = \\ \delta_{IJ}\delta_{MM'}E_I + \mel{\Psi_I^M}{\hat{H}_{SOC}+\hat{H}_{SSC}}{\Psi_J^{M'}} \\ \end{multline} 
where $I$ and $J$ label the CASSCF roots, while $M$ and $M'$ denote the spin sublevels. Although the spin-sublevel dependence is not explicit in the spin-free CASSCF roots, it is introduced through Clebsch-Gordan coefficients according to the Wigner-Eckart theorem~\cite{NeeseSolomon1998,GanyushinNeese2013_CASSOC}. The SOC matrix elements were evaluated using the spin-orbit mean-field (SOMF) approximation~\cite{Hess1996_SOMF,Neese2005_SOMF}. Finally, $E_I^{\mathrm{CASSCF}}$ denotes the spin-free electronic energy of the $I$th CASSCF root.

Diagonalization of the QDPT matrix yields the spin-orbit coupled eigenstates and the corresponding relativistic energy levels. The resulting multiplet structure was assigned to effective $J$ quantum numbers by inspecting the quasi-degeneracies in the energy spectrum, as described in SI, section~S1.5.

For all CASSCF-QDPT calculations, the def2-SVP basis set was used together with the corresponding effective core potential (ECP) for Er~\cite{WeigendAhlrichs2005_Def2,CaoDolg2001_LnECPBasis,CaoDolg2002_LnSegmented}. Density fitting approximations were applied to accelerate the calculations, using the def2/J and def2-SVP/C auxiliary basis sets in the Coulomb and correlation auxiliary-basis slots, respectively~\cite{Weigend2006_Def2J,Hellweg2007_AuxC,ChmelaHarding2018_LnAuxC}.

\section*{\large Data availability}

The main data supporting the findings of this study are available in the paper and its Supplementary Information. Further numerical data are available from the authors upon reasonable request.

\section*{\large Acknowledgements}

G.D. is grateful to Laura Dolecsekné Mócsán for her assistance in constructing the impurity-doped talc structures, selecting the color scheme, and preparing the geometric illustrations. Z.B. acknowledges the financial support of the János Bolyai Research Fellowship of the Hungarian Academy of Sciences. We acknowledge the support of the European Union under Horizon Europe for the QRC-4-ESP project (Grant Agreement 101129663) and the QUEST project (Grant Agreement 101156088). N.T.S. acknowledges the EU project QuSPARC (Grant No. 101186889), the Vinnova projects (2025-03848, and QUASIC within QSIP, grant 2024-03597), and the EU via the Swedish Agency for Economic and Regional Growth (20370271).
The computations were enabled by resources provided by the National Academic Infrastructure for Supercomputing in Sweden (NAISS) and the Swedish National Infrastructure for Computing (SNIC) at NSC, partially funded by the Swedish Research Council through grant agreements no. 2022-06725 and no. 2018-05973.
We acknowledge the Digital Government Development and Project Management Ltd. for awarding us access to the Komondor HPC facility based in Hungary

\section*{\large Competing interests}

The authors declare no competing interests.

\section*{\large Author contributions}

G.D. and Z.B. carried out the DFT and WFT calculations, respectively. G.D., Z.B., N.T.S. and V.I. analysed the data.  G.D., Z.B., and V.I. wrote the manuscript with inputs from all coauthors. The work was supervised by  Z.B. and V.I. 

\newpage

\bibliography{references}

\end{document}

\title{Supplementary Information \\ for \\  Natural van der Waals silicates as hosts for telecom quantum emitters: the case of erbium-doped talc}
\date{\today}

\author{Gell\'{e}rt Dolecsek}
\affiliation{Department of Physics of Complex Systems, Eötvös Loránd University, Egyetem tér 1-3, H-1053 Budapest, Hungary}
\affiliation{MTA–ELTE Lend\"{u}let "Momentum" NewQubit Research Group, Pázmány Péter, Sétány 1/A, 1117 Budapest, Hungary}

\author{Zsolt Benedek}
\affiliation{Department of Physics of Complex Systems, Eötvös Loránd University, Egyetem tér 1-3, H-1053 Budapest, Hungary}
\affiliation{MTA–ELTE Lend\"{u}let "Momentum" NewQubit Research Group, Pázmány Péter, Sétány 1/A, 1117 Budapest, Hungary}

\author{Nguyen Tien Son}
\affiliation{Department of Physics, Chemistry and Biology, Link\"oping University, SE-581 83 Link\"oping, Sweden}

\author{Viktor Ivády}
\email{ivady.viktor@ttk.elte.hu}
\affiliation{Department of Physics of Complex Systems, Eötvös Loránd University, Egyetem tér 1-3, H-1053 Budapest, Hungary}
\affiliation{MTA–ELTE Lend\"{u}let "Momentum" NewQubit Research Group, Pázmány Péter, Sétány 1/A, 1117 Budapest, Hungary}

\maketitle

\tableofcontents

%\section{Details and convergence of GW+BSE calculations}

%\bibliographystyle{plain}

\newpage

\section{Supplementary results}
\label{results}

\subsection{Comparison of R2SCAN and HSE06 functionals}
\label{subsec:functional_test}

In order to evaluate the performance of the R2SCAN functional relative to the more expensive hybrid alternative, HSE06, we ran comparative test calculations. In particular, the formation energy ($E^f$) and M-O bond lengths (where M stands for Mg or Er) were determined using identical VASP settings apart from functional choice. The results are summarized in Tab.~\ref{tab:comparison}. Overall, formation energies (herein omitting the $E_corr$ term) agree within an error range of $1$~eV, while bond distances are essentially identical.    

\begin{table}[h!]
\centering
\caption{Comparison of R2SCAN and HSE06 results.}
\label{tab:comparison}

\begin{tabular}{|l|cc|cc|}
\hline
\textbf{Defect} &
\multicolumn{2}{c|}{$E^{f}$ (eV)} &
\multicolumn{2}{c|}{$r(\mathrm{M{-}O})$ (\AA)} \\
\cline{2-5}
 & \textbf{R2SCAN} & \textbf{HSE06} & \textbf{R2SCAN} & \textbf{HSE06} \\
\hline
Bulk talc ($Mg_1$) & - & - & 2.05-2.06 & 2.04-2.07 \\
Bulk talc ($Mg_2$) & - & - & 2.04-2.07 & 2.05-2.08 \\
\hline
$\mathrm{Er}_{\mathrm{Mg1}}$ & (optimization failed) & -1.5 & (optimization failed) & 2.29-2.31 \\
$\mathrm{Er}_{\mathrm{Mg2}}$ & -1.6 & -2.4 & 2.20-2.29 & 2.20-2.29\\
\hline
$V_{\mathrm{H}}\mathrm{Er}_{\mathrm{Mg1}}$ & -3.3 & -4.3 & 2.02-2.34& 2.02-2.33\\
$V_{\mathrm{H}}\mathrm{Er}_{\mathrm{Mg2}}$ & -3.3 & -4.3 & 2.02-2.33& 2.02-2.33\\
\hline
$2V_{\mathrm{H}}\mathrm{Er}_{\mathrm{Mg1}}$ & -0.8 & -1.4 & 2.04-2.22 & 2.04-2.22 \\
$2V_{\mathrm{H}}\mathrm{Er}_{\mathrm{Mg2}}$ & -0.3 & -1.2 & 2.06-2.25 & 2.09-2.24 \\
\hline
\end{tabular}
\end{table}

\subsection{Interstitial erbium defects in talc}
\label{subsec:interlayer}

\begin{figure*}
\includegraphics[width=0.8\textwidth]{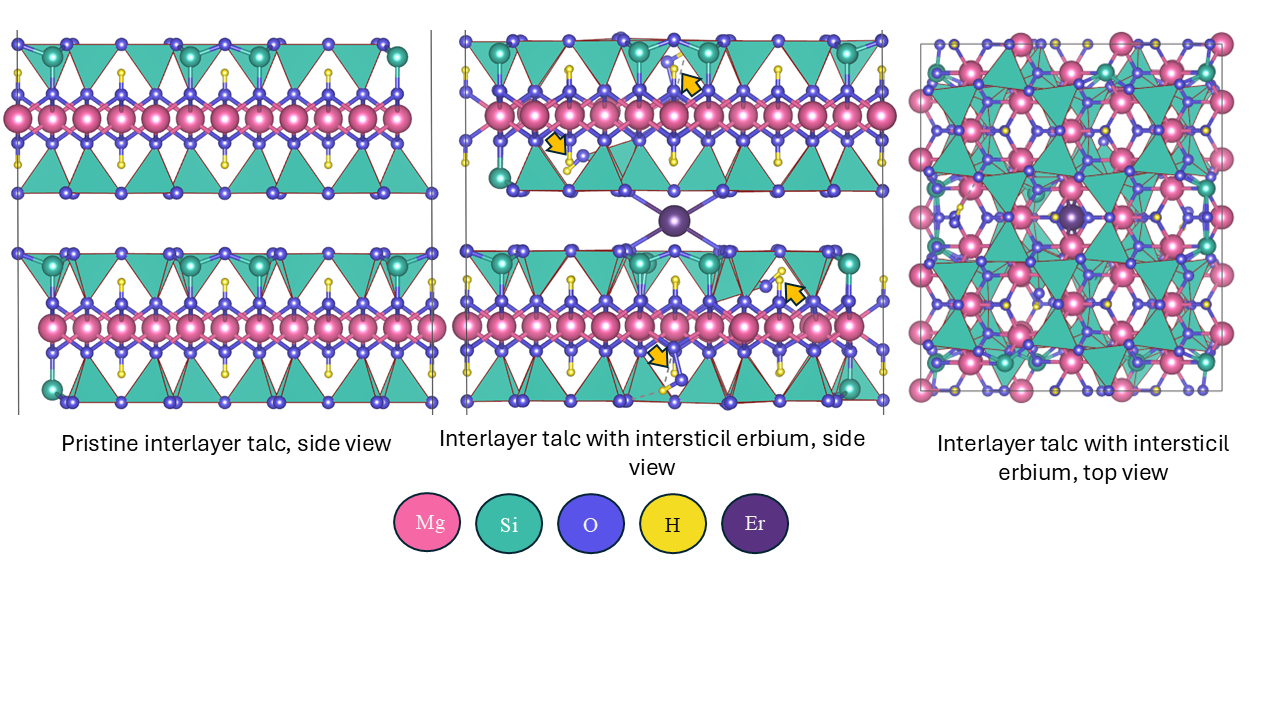}
	\caption{The structure of a two layer pristine talc and with an interstitial Er atom. Purple, turquoise, mauve, blue, and yellow spheres indicate the positions of erbium, silicon, magnesium, oxygen, and hydrogen atoms.}
	\label{fig:geo_interlayer}  
\end{figure*}

In this section, we investigate bilayer talc represented by a 3x2x2 supercell. A positively charged erbium ion is introduced between the layers to examine whether erbium preferentially incorporates into the talc lattice (as described in the main text) or tends to occupy an interstitial position, see \ref{fig:geo_interlayer}. 
 
The presence of Er$^{3+}$ does not alter the interlayer oxygen-oxygen distances significantly (pristine: $4.27\text{\AA}$; with Er$^{3+}$: $4.16\text{\AA}$), however, significant changes in the orientation of nearby OH groups can be observed, see yellow arrows in Fig.\ref{fig:geo_interlayer}. 

\begin{figure*}
\includegraphics[width=0.8\textwidth]{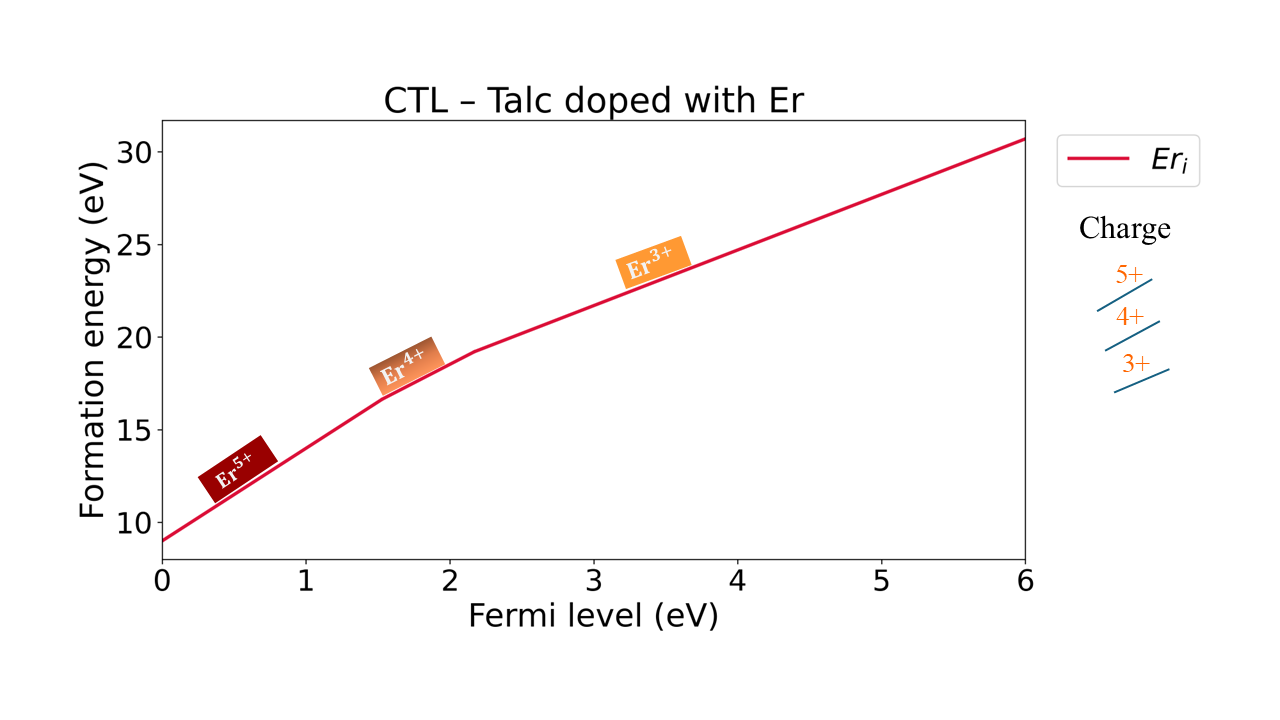}
	\caption{Formation energy of the interstitial erbium centers in talc as a function of the Fermi level.}
	\label{fig:formation_energy}  
\end{figure*}

The calculated formation energy of interstitial Er$^{n+}$ (n=3-5 in the relevant Fermi level range) is high: $\text{9.0}$ eV is calculated even at the most favorable p-type conditions (Fig.~\ref{fig:formation_energy}), indicating that erbium preferentially occupies the Mg$^{2+}$ sites.

\subsection{Construction of cluster models}
\label{subsec:cluster}

\begin{figure*}[h]
\includegraphics[width=\textwidth]{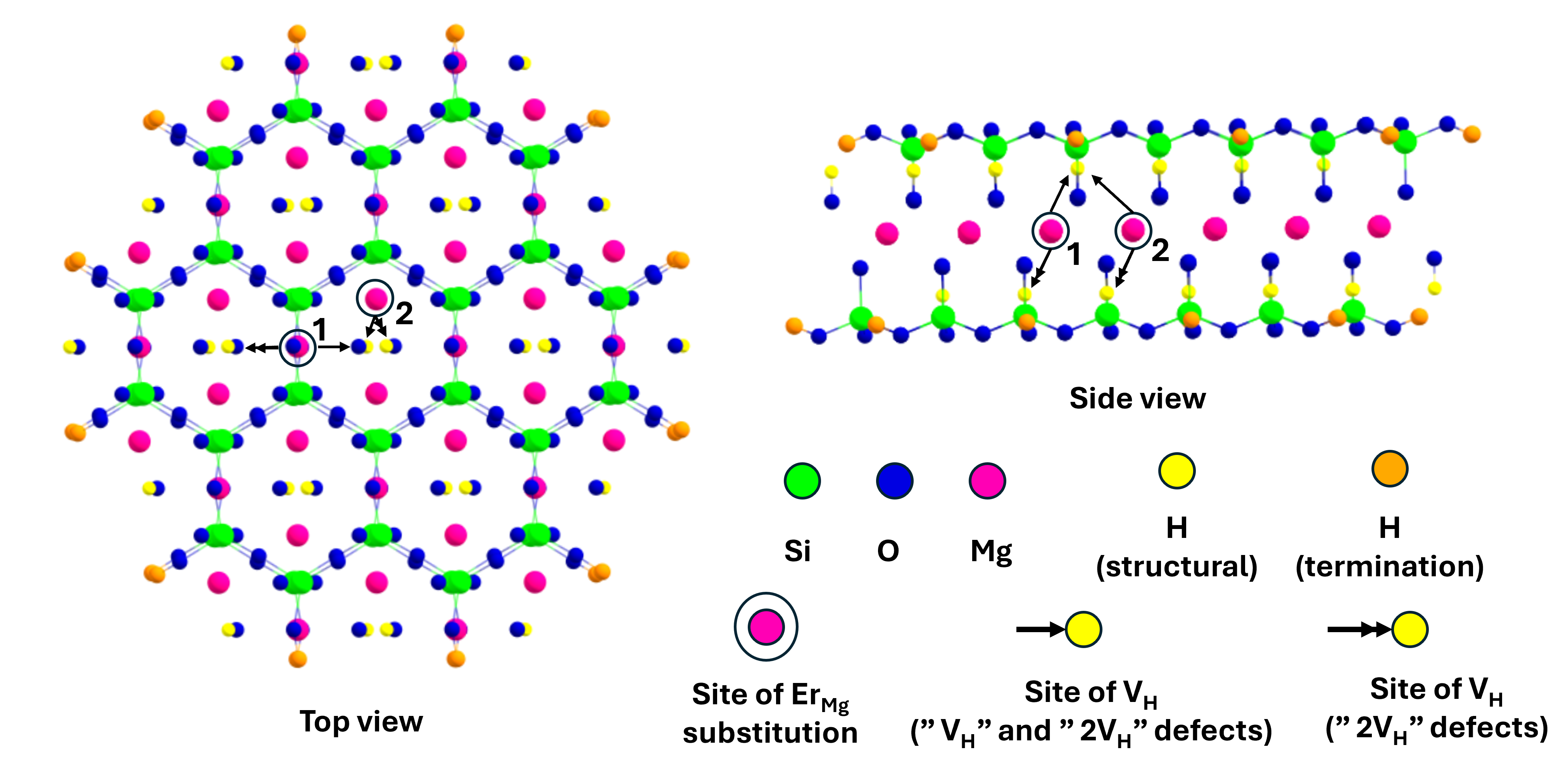}
	\caption{Visualization of the 298-atom cluster model used to study the erbium-containing defects in talc.}
	\label{fig:cluster_model}  
\end{figure*}

The cluster model for each species of Tab.~\ref{main-tab:nomenclature} was prepared as follows. The starting point is the DFT-optimized supercell geometry (specific to the given defect), from which the position of $274-n_{V_H}$ atoms were cut, where $n_{V_H}$ is the number of hydrogen vacancy sites ($V_H$) in the studied defect. The 274 atoms would correspond to Si, O, Mg, and yellow-labeled structural H atoms, as shown in Fig.~\ref{fig:cluster_model}, in the case of bulk talc; for a given erbium-containing defect, however, an Er atom is selected in lieu of Mg at site "1" or "2" (for $Er_{Mg1}$ and $Er_{Mg2}$ substitutionals, respectively), and one or two hydrogen atoms from the OH group(s) closest to Er is absent for "$V_H$" and "2$V_H$" type defects, respectively. Finally, 24 terminal H atoms (labeled by orange in Fig. S1) were added to the outer oxygen atoms of the two tetrahedral silicate sheets, at which the Si-O bonds were cut to create the finite-size cluster. These terminal atoms were placed at the position of the missing Si neighbors, but the O-H distance was reduced to 1.01 $\AA$, without changing angles or dihedral angles.

The above described process of cluster model construction was motivated by the work of Tunega and Turi Nagy~\cite{Tunega1993TalcElectronStructure}, where the 298-atom cluster of Fig. S1 was peresented as a charge-neutral molecular model for bulk talc. It was also demonstrated that the electronic properties of the central part of the cluster (i.e. the vicinity of centers "1" and "2") closely resemble those of infinite monolayer talc. 

\subsection{Comparison of supercell and cluster models }
\label{subsec:LOCPOT}

To validate our cluster model, we assessed whether it is sufficiently large to reproduce the electrostatic environment of the embedded Er center in the investigated systems. For this purpose, we compared the coordinate-dependence of the local potential energy of electrons near Er, as obtained from two periodic VASP calculations, i.e. (1) cluster model within a 25 $\AA$ vacuum box and (2) the corresponding reference supercell model, as described under Methods in the main text. Both calculations were performed using the R2SCAN functional, with identical settings.

\begin{figure*}[h]
\includegraphics[width=\textwidth]{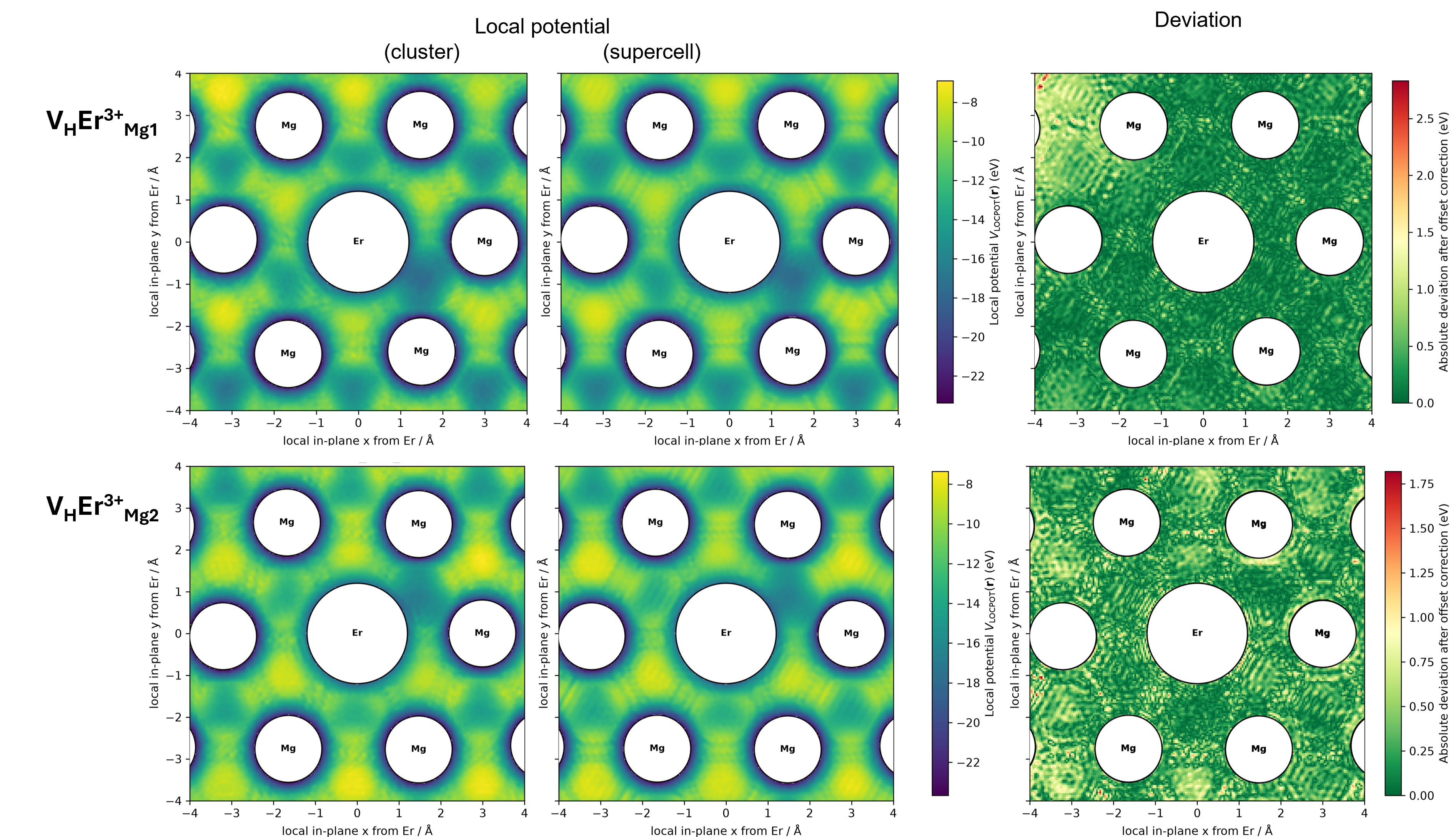}
	\caption{Comparison of local potential energies of electrons in cluster models (left column) and supercells (middle column) in the vicinity of the Er center. Plots of the right column indicate coordinate-dependent deviations.}
	\label{fig:LOCPOT}  
\end{figure*}

The VASP \texttt{LOCPOT} file stores the local potential on a uniform real-space grid, which is to be compared for the two models. Herein, this comparison is provided for the plane of Er and Mg atoms, in the vicinity of Er (Fig.~\ref{fig:LOCPOT}).      

Since the absolute zero of the potential is set independently in the calculations, the two
\texttt{LOCPOT} maps were compared only after removal of the constant offset, following
the usual potential-alignment logic employed in supercell defect calculations
\cite{Freysoldt2014RMP}. The latter was determined by minimizing the mean absolute
difference between the local potentials. Furthermore, near-core regions (shown in white in
Fig.~\ref{fig:LOCPOT}) were excluded from both the visual comparison and all numerical
diagnostics, as the local potential close to the nuclei is dominated by steep atomic-scale
features, the frozen-core/projector-augmented-wave representation, and grid interpolation
details rather than by the varying embedding field of interest
\cite{blochl1994paw,Kresse1999PAW}. Specifically, circular masks of radius
$1.2~\mathrm{\AA}$ around Er and $0.8~\mathrm{\AA}$ around Mg were applied to conservatively
exclude the PAW near-core regions from the comparison.

A cluster model that is too small is expected to show artificial boundary effects in the potential map, because the electrostatic response of the real periodic environment is replaced by vacuum and by the finite termination of the cluster. In practice this may appear as an unphysical long-range gradient or curvature of the offset-corrected \texttt{LOCPOT} toward the edges of the plotted region, even if the immediate coordination shell is geometrically identical. Accordingly, as the measure of cluster model quality, we plot the deviation between the two offset-corrected potentials over the chemically relevant region (Fig.~\ref{fig:LOCPOT}, right column). In Tab.~\ref{tab:LOCPOT}, we also report a scalar similarity index ($S_{LOCPOT}$), computed from the mean absolute error of local potential energies as
\begin{equation}
S_\mathrm{LOCPOT}
=
1-
\frac{
\left\langle
\left|
V_{cluster}(x,y) - V_{supercell}(x,y)
\right|
\right\rangle
}{
\max(V_{cluster},V_{supercell})
-
\min(V_{cluster},V_{supercell})}
\end{equation}

The value of $S_{LOCPOT}$ approaches 1 in the case of essentially indentcal potential energy maps, while reduces nearly to zero if the model-related variation of data becomes commensurable to the maximal potential energy difference observed within one \texttt{LOCPOT} map.

Fig.~\ref{fig:LOCPOT} and Tab.~\ref{tab:LOCPOT} clearly indicate that the local potential energy maps sampled from cluster models properly reflect those sampled from corresponding supercells. No visually detectable effect of cluster termination can be observed, and all $S_{LOCPOT}$ values remain above 0.98, showing perfect similarity.

\begin{table}[h!]
\centering
\caption{Similarity index ($S_{LOCPOT}$) calculated from cluster-model and supercell-model local potential energy maps. }
\label{tab:LOCPOT}

\begin{tabular}{|l|c|}
\hline
Defect & $S_{LOCPOT}$ \\
\hline
$V_HEr^{3+}_{Mg1}$ & 0.982 \\
$V_HEr^{3+}_{Mg2}$ & 0.982 \\
\hline
\end{tabular}
\end{table}

\subsection{Boltzmann statistics of $^4I_{13/2} \rightarrow ^4I_{15/2}$ and $^4I_{7} \rightarrow ^4I_{8}$ emissions }
\label{subsec:Boltzmann}

The radiative photoluminescence (PL) rates were obtained by post-processing the SOC-corrected ORCA one-photon spectroscopy output by an in-house developed script. In the first step, two parts of the ORCA output are parsed: the SOC-corrected combined electric-dipole/magnetic-dipole/electric-quadrupole absorption table and the ORCA Boltzmann population block. Each row of the ORCA spectrum has the form $
f \rightarrow i $
where (f) is the final state of PL, i.e. the lower-energy state and (i) is the initial state of PL, i.e. the upper-energy state. The same transition is interpreted as an emission channel in the reverse direction ($
i\rightarrow f $). 
Only transitions for which the $f$ state  belongs to the selected lower manifold and $i$ state belongs to the selected emitting manifold are retained. For $Er^{3+}$ the default lower manifold is taken as states (0,$\ldots$,15), corresponding to the ($^{4}I_{15/2}$) multiplet, and the default emitting manifold is taken as states (16,$\ldots$,29), corresponding to the ($^{4}I_{13/2}$) multiplet. For $Er^{4+}$, the corresponding default windows are states (0,$\ldots$,16) for the ($^{5}I_{8}$) lower manifold and states (17,$\ldots$,31) for the ($^{5}I_{7}$) emitting manifold. The transition energy ($\Delta E_{if}$), as well as magnetic and electric dipole induced oscillator strengths ($f_{if}^{\mathrm{M}}$,$f_{if}^{\mathrm{E}}$) are extracted for all individual $
i\rightarrow f $ transitions separately. We note here that oscillator strengths printed by ORCA are population-weighted absorption oscillator strengths. Therefore, before using them to calculate emission rates,the population weighing needs to be removed. We further note that electric and magnetic transition dipole moments ($\mu_M$, $\mu_B$) can be obtained from the oscillator strength using the general formula of
\begin{equation}
\mu_{if} = \left(
\frac{3 f_{if}}{2\Delta E_{if}}
\right)^{1/2}   
\end{equation}
The electric-dipole and magnetic-dipole spontaneous-emission rates are then computed from the corresponding oscillator strengths as
\begin{equation}
k_{if}^{\mathrm{M}}
=
\frac{2\alpha^3}{t_{\mathrm{au}}}\,
n^3\,
\left(\Delta E_{if}\right)^2
f_{if}^{\mathrm{M}} .
\end{equation}
and
\begin{equation}
k_{if}^{\mathrm{E}}
=
\frac{2\alpha^3}{t_{\mathrm{au}}}\,
n\,
\left(\Delta E_{if}\right)^2
f_{if}^{\mathrm{E}} .
\end{equation}
where $n$ is the refractive index of the host medium, $t_{\mathrm{au}}$ is the atomic unit of time, and $\alpha$ is the fine-structure constant.Thus, the we apply an $n$-scaling to the electric-dipole contribution and an $n^3$-scaling to the magnetic-dipole contribution. Finally, the rate of a given individual Stark transition is obtained as
\begin{equation}
k_{if}
= \frac{1}{\tau_{if}} =
k_{if}^{\mathrm{M}} + k_{if}^{\mathrm{E}}
\end{equation}

From the point of view of quantum applications, these individual rates  determine the brightness of a selected, well-defined optical transition that can be enhanced by the Purcell effect. Therefore, the highest obtained $k_{if}$ values are provided in Tab.~\ref{main-tab:PL_Q} of the main text.

General optoelectronic applications, on the other hand, do not require single-photon emission and can utilize all Stark transitions. In this case, the brightness is related to the total rate ($k^{total}$) of all available ZPLs. To investigate this total rate, we prepared a Boltzmann averaging as follows.

When investigating overall brightness from multiple ZPLs, $k_{if}$ is not equal to the expected rate of photon emission, as the relative Boltzmann population of the initial Stark state ($0<p_i<1$) also has to be accounted for by a prefactor. In the case of non-Kramers ions, such as $Er^{4+}$, the radiative rate belonging to individual ZPLs ($k_{if}^{eff}$) corresponds simply to 

\begin{equation}
k_{if}^{eff}
= \frac{1}{\tau_{if}^{eff}} =
p_ik_{if}
\end{equation}

For Kramers ions, such as $Er^{3+}$, however, a single ZPL arises from four individual rates, as two initial ($i1,i2$) and two final ($f1,f2$) states are degenarate. Therefore, the PL rate observable in experiments ($k_{if}^{eff, Kramers}$) is

\begin{equation}
k_{if}^{eff, Kramers}
= \frac{1}{\tau_{if}^{eff, Kramers}} =
\sum_{i=1,2}\sum_{f=1,2}p_ik_{if}
\end{equation}  

 For a given emitting Stark state (i), the total electric- and magnetic-dipole rates are obtained by summing over all selected lower-manifold final states:

\begin{equation}
k_i^{\mathrm{E}}= \sum_{f\in \mathcal{F}}k_{if}^{\mathrm{E}}
\end{equation}

\begin{equation}
k_i^{\mathrm{M}}= \sum_{f\in \mathcal{F}}k_{if}^{\mathrm{M}}
\end{equation}

\begin{equation}
k_{i}=
k_{i}^{\mathrm{M}} + k_{i}^{\mathrm{E}}
\end{equation}

Here, $\mathcal{F}$ denotes the selected final-state manifold. The thermal averaging is then performed on the initial side of PL, as also indicated in equations (3) and (4) in the main text:

\begin{equation}
k^{total} = \sum_ip_ik_i
\end{equation}

For the compact characterization of the overall transition strength (which is presented in Tab.~\ref{main-tab:PL} of the main text), we define a representative emission energy as the Boltzmann-weighted average over all retained emission lines,

\begin{equation}
\Delta E_{\mathrm{rep}}
=
\frac{
\sum_i p_i \sum_{f\in\mathcal{F}} \Delta E_{if}
}{
\sum_i p_i \sum_{f\in\mathcal{F}} 1
} .
\end{equation}

Using this representative energy, effective electric and magnetic transition moments are back-calculated from the thermally averaged rates. The effective electric transition dipole is obtained as

\begin{equation}
\mu_{\mathrm{E}}^{\mathrm{eff}}
=
\left[
\frac{
3 t_{\mathrm{au}} \sum_i p_i k_i^{\mathrm{E}}
}{
4\alpha^3 n \left(\Delta E_{\mathrm{rep}}\right)^3
}
\right]^{1/2} .
\end{equation}
and is reported in Debye. The effective magnetic transition dipole is obtained as

\begin{equation}
\mu_{\mathrm{M}}^{\mathrm{eff}}
=
\left[
\frac{
3 t_{\mathrm{au}} \sum_i p_i k_i^{\mathrm{M}}
}{
4\alpha^3 n^3 \left(\Delta E_{\mathrm{rep}}\right)^3
}
\right]^{1/2} .
\end{equation}

and is reported in units of ($\mu_{\mathrm{B}}$).

Finally, we provide a characteristic emission-energy interval for all defects, which was determined from the transition-resolved PL
intensity. For each
$i \rightarrow f$ transition, the intensity can be defined as $k^{eff}$, and one can sort all transitions by their
emission energies. Based on the gathered data, we report the narrowest contiguous energy interval for which
the summed intensity reaches at least \(95\%\) of the total PL intensity.

We note that due to the limitations of the QDPT treatment, theoretically exact degeneracies might break in numerical results. This effect, however, has negligible influence on the rates that determine practical applications.

% Preamble packages needed:
% \usepackage{pdflscape}
% \usepackage{adjustbox}
% \usepackage{booktabs}
% \usepackage{makecell}

\newcommand{\initkp}[1]{$^4I_{13/2}$ KP~#1}
\newcommand{\finalkp}[1]{$^4I_{15/2}$ KP~#1}

% Detailed Kramers-pair-resolved PL rates
\begin{table}[t]
\centering
\tiny
\setlength{\tabcolsep}{1.6pt}
\renewcommand{\arraystretch}{0.82}
\caption{Detailed Kramers-pair-resolved PL rates for \textbf{free $Er^{3+}$ ion}. "KP" denotes Kramers pair.}
\label{tab:pl-transition-resolved}
\begin{adjustbox}{max width=\linewidth, max totalheight=0.90\textheight, keepaspectratio}
\begin{tabular}{@{}|l|l|c|c|c|c|c|c|c|c|c|c|@{}}
\toprule
\makecell[c]{Initial\\state} & \makecell[c]{Final\\state} & \makecell[c]{$\Delta E$\\(eV)} & \makecell[c]{$\mu_\mathrm{M}$\\($\mu_\mathrm{B}$)} & \makecell[c]{$\mu_\mathrm{E}$\\(D)} & \makecell[c]{$k_\mathrm{if}$\\(s$^{-1}$)} & \makecell[c]{$\tau_{if}$\\(ms)} & $p_i$ & \makecell[c]{$ k_\mathrm{if}^{eff, Kramers}$\\(s$^{-1}$)} & \makecell[c]{$\tau^{eff, Kramers}$\\(ms)} & \makecell[c]{E\\(\%)} & \makecell[c]{M\\(\%)} \\
\midrule
\initkp{1} & \finalkp{1} & 0.75 & $1.00\times10^{0}$ & $0$ & $6.03\times10^{0}$ & $1.66\times10^{2}$ & $1.43\times10^{-1}$ & $8.61\times10^{-1}$ & $1.16\times10^{3}$ & 0.0 & 100.0 \\
\initkp{1} & \finalkp{2} & 0.75 & $5.10\times10^{-1}$ & $0$ & $1.57\times10^{0}$ & $6.37\times10^{2}$ & $1.43\times10^{-1}$ & $2.24\times10^{-1}$ & $4.46\times10^{3}$ & 0.0 & 100.0 \\
\initkp{1} & \finalkp{3} & 0.75 & $2.52\times10^{-1}$ & $0$ & $3.85\times10^{-1}$ & $2.60\times10^{3}$ & $1.43\times10^{-1}$ & $5.50\times10^{-2}$ & $1.82\times10^{4}$ & 0.0 & 100.0 \\
\initkp{1} & \finalkp{4} & 0.75 & $1.41\times10^{-1}$ & $0$ & $1.20\times10^{-1}$ & $8.34\times10^{3}$ & $1.43\times10^{-1}$ & $1.71\times10^{-2}$ & $5.84\times10^{4}$ & 0.0 & 100.0 \\
\initkp{1} & \finalkp{5} & 0.75 & $1.44\times10^{-2}$ & $0$ & $1.25\times10^{-3}$ & $8.00\times10^{5}$ & $1.43\times10^{-1}$ & $1.79\times10^{-4}$ & $5.60\times10^{6}$ & 0.0 & 100.0 \\
\initkp{1} & \finalkp{6} & 0.75 & $1.65\times10^{-1}$ & $0$ & $1.64\times10^{-1}$ & $6.08\times10^{3}$ & $1.43\times10^{-1}$ & $2.35\times10^{-2}$ & $4.26\times10^{4}$ & 0.0 & 100.0 \\
\initkp{1} & \finalkp{7} & 0.75 & $3.52\times10^{-2}$ & $0$ & $7.46\times10^{-3}$ & $1.34\times10^{5}$ & $1.43\times10^{-1}$ & $1.07\times10^{-3}$ & $9.38\times10^{5}$ & 0.0 & 100.0 \\
\bottomrule
\end{tabular}
\end{adjustbox}
\end{table}

% Detailed Kramers-pair-resolved PL rates
\begin{table}
\centering
\tiny
\setlength{\tabcolsep}{1.6pt}
\renewcommand{\arraystretch}{0.82}
\caption{Detailed Kramers-pair-resolved PL rates for \textbf{$Er^{3+}_{Mg1}$}. "KP" denotes Kramers pair.}
\label{tab:pl-transition-resolved}
\begin{adjustbox}{max width=\linewidth, max totalheight=0.90\textheight, keepaspectratio}
\begin{tabular}{@{}|l|l|c|c|c|c|c|c|c|c|c|c|@{}}
\toprule
\makecell[c]{Initial\\state} & \makecell[c]{Final\\state} & \makecell[c]{$\Delta E$\\(eV)} & \makecell[c]{$\mu_\mathrm{M}$\\($\mu_\mathrm{B}$)} & \makecell[c]{$\mu_\mathrm{E}$\\(D)} & \makecell[c]{$k_\mathrm{if}$\\(s$^{-1}$)} & \makecell[c]{$\tau_{if}$\\(ms)} & $p_i$ & \makecell[c]{$ k_\mathrm{if}^{eff, Kramers}$\\(s$^{-1}$)} & \makecell[c]{$\tau^{eff, Kramers}$\\(ms)} & \makecell[c]{E\\(\%)} & \makecell[c]{M\\(\%)} \\
\midrule
\initkp{1} & \finalkp{1} & 0.76 & $4.71\times10^{-1}$ & $2.63\times10^{-4}$ & $5.07\times10^{0}$ & $1.97\times10^{2}$ & $2.43\times10^{-1}$ & $1.23\times10^{0}$ & $8.11\times10^{2}$ & 0.2 & 99.8 \\
\initkp{1} & \finalkp{2} & 0.75 & $3.03\times10^{-1}$ & $6.25\times10^{-5}$ & $2.06\times10^{0}$ & $4.85\times10^{2}$ & $2.43\times10^{-1}$ & $5.01\times10^{-1}$ & $2.00\times10^{3}$ & 0.0 & 100.0 \\
\initkp{1} & \finalkp{3} & 0.75 & $9.13\times10^{-1}$ & $1.64\times10^{-4}$ & $1.84\times10^{1}$ & 54.438 & $2.43\times10^{-1}$ & $4.47\times10^{0}$ & $2.24\times10^{2}$ & 0.0 & 100.0 \\
\initkp{1} & \finalkp{4} & 0.74 & $3.74\times10^{-1}$ & $1.42\times10^{-4}$ & $3.02\times10^{0}$ & $3.31\times10^{2}$ & $2.43\times10^{-1}$ & $7.35\times10^{-1}$ & $1.36\times10^{3}$ & 0.1 & 99.9 \\
\initkp{1} & \finalkp{5} & 0.74 & $2.94\times10^{-1}$ & $2.52\times10^{-4}$ & $1.86\times10^{0}$ & $5.38\times10^{2}$ & $2.43\times10^{-1}$ & $4.52\times10^{-1}$ & $2.21\times10^{3}$ & 0.4 & 99.6 \\
\initkp{1} & \finalkp{6} & 0.71 & $7.35\times10^{-2}$ & $9.11\times10^{-5}$ & $1.01\times10^{-1}$ & $9.85\times10^{3}$ & $2.43\times10^{-1}$ & $2.47\times10^{-2}$ & $4.05\times10^{4}$ & 0.7 & 99.3 \\
\initkp{1} & \finalkp{7} & 0.70 & $4.79\times10^{-2}$ & $1.05\times10^{-4}$ & $4.24\times10^{-2}$ & $2.36\times10^{4}$ & $2.43\times10^{-1}$ & $1.03\times10^{-2}$ & $9.71\times10^{4}$ & 2.3 & 97.7 \\
\bottomrule
\end{tabular}
\end{adjustbox}
\end{table}

\newpage

% Detailed Kramers-pair-resolved PL rates
\begin{table}
\centering
\tiny
\setlength{\tabcolsep}{1.6pt}
\renewcommand{\arraystretch}{0.82}
\caption{Detailed Kramers-pair-resolved PL rates for \textbf{$Er^{3+}_{Mg2}$}. "KP" denotes Kramers pair.}
\label{tab:pl-transition-resolved}
\begin{adjustbox}{max width=\linewidth, max totalheight=0.90\textheight, keepaspectratio}
\begin{tabular}{@{}|l|l|c|c|c|c|c|c|c|c|c|c|@{}}
\toprule
\makecell[c]{Initial\\state} & \makecell[c]{Final\\state} & \makecell[c]{$\Delta E$\\(eV)} & \makecell[c]{$\mu_\mathrm{M}$\\($\mu_\mathrm{B}$)} & \makecell[c]{$\mu_\mathrm{E}$\\(D)} & \makecell[c]{$k_\mathrm{if}$\\(s$^{-1}$)} & \makecell[c]{$\tau_{if}$\\(ms)} & $p_i$ & \makecell[c]{$ k_\mathrm{if}^{eff, Kramers}$\\(s$^{-1}$)} & \makecell[c]{$\tau^{eff, Kramers}$\\(ms)} & \makecell[c]{E\\(\%)} & \makecell[c]{M\\(\%)} \\
\midrule
\initkp{1} & \finalkp{1} & 0.76 & $4.54\times10^{-1}$ & $4.66\times10^{-4}$ & $4.72\times10^{0}$ & $2.12\times10^{2}$ & $2.42\times10^{-1}$ & $1.15\times10^{0}$ & $8.73\times10^{2}$ & 0.5 & 99.5 \\
\initkp{1} & \finalkp{2} & 0.75 & $2.81\times10^{-1}$ & $1.41\times10^{-4}$ & $1.78\times10^{0}$ & $5.62\times10^{2}$ & $2.42\times10^{-1}$ & $4.31\times10^{-1}$ & $2.32\times10^{3}$ & 0.1 & 99.9 \\
\initkp{1} & \finalkp{3} & 0.75 & $9.33\times10^{-1}$ & $1.82\times10^{-4}$ & $1.92\times10^{1}$ & 52.202 & $2.42\times10^{-1}$ & $4.65\times10^{0}$ & $2.15\times10^{2}$ & 0.0 & 100.0 \\
\initkp{1} & \finalkp{4} & 0.74 & $3.77\times10^{-1}$ & $9.80\times10^{-5}$ & $3.07\times10^{0}$ & $3.26\times10^{2}$ & $2.42\times10^{-1}$ & $7.45\times10^{-1}$ & $1.34\times10^{3}$ & 0.0 & 100.0 \\
\initkp{1} & \finalkp{5} & 0.74 & $2.72\times10^{-1}$ & $8.15\times10^{-5}$ & $1.59\times10^{0}$ & $6.30\times10^{2}$ & $2.42\times10^{-1}$ & $3.85\times10^{-1}$ & $2.60\times10^{3}$ & 0.0 & 100.0 \\
\initkp{1} & \finalkp{6} & 0.71 & $7.00\times10^{-2}$ & $4.01\times10^{-4}$ & $1.06\times10^{-1}$ & $9.42\times10^{3}$ & $2.42\times10^{-1}$ & $2.58\times10^{-2}$ & $3.88\times10^{4}$ & 13.7 & 86.3 \\
\initkp{1} & \finalkp{7} & 0.70 & $4.15\times10^{-2}$ & $4.45\times10^{-4}$ & $4.87\times10^{-2}$ & $2.05\times10^{4}$ & $2.42\times10^{-1}$ & $1.18\times10^{-2}$ & $8.47\times10^{4}$ & 35.7 & 64.3 \\
\bottomrule
\end{tabular}
\end{adjustbox}
\end{table}

% Detailed Kramers-pair-resolved PL rates
\begin{table}
\centering
\tiny
\setlength{\tabcolsep}{1.6pt}
\renewcommand{\arraystretch}{0.82}
\caption{Detailed Kramers-pair-resolved PL rates for $V_HEr^{3+}_{Mg1}$. "KP" denotes Kramers pair.}
\label{tab:pl-transition-resolved}
\begin{adjustbox}{max width=\linewidth, max totalheight=0.90\textheight, keepaspectratio}
\begin{tabular}{@{}|l|l|c|c|c|c|c|c|c|c|c|c|@{}}
\toprule
\makecell[c]{Initial\\state} & \makecell[c]{Final\\state} & \makecell[c]{$\Delta E$\\(eV)} & \makecell[c]{$\mu_\mathrm{M}$\\($\mu_\mathrm{B}$)} & \makecell[c]{$\mu_\mathrm{E}$\\(D)} & \makecell[c]{$k_\mathrm{if}$\\(s$^{-1}$)} & \makecell[c]{$\tau_{if}$\\(ms)} & $p_i$ & \makecell[c]{$ k_\mathrm{if}^{eff, Kramers}$\\(s$^{-1}$)} & \makecell[c]{$\tau^{eff, Kramers}$\\(ms)} & \makecell[c]{E\\(\%)} & \makecell[c]{M\\(\%)} \\
\midrule
\initkp{1} & \finalkp{1} & 0.75 & $8.98\times10^{-1}$ & $3.00\times10^{-3}$ & $1.92\times10^{1}$ & 51.996 & $2.66\times10^{-1}$ & $5.11\times10^{0}$ & $1.96\times10^{2}$ & 5.1 & 94.9 \\
\initkp{1} & \finalkp{2} & 0.75 & $3.17\times10^{-1}$ & $1.65\times10^{-3}$ & $2.53\times10^{0}$ & $3.95\times10^{2}$ & $2.66\times10^{-1}$ & $6.72\times10^{-1}$ & $1.49\times10^{3}$ & 11.6 & 88.4 \\
\initkp{1} & \finalkp{3} & 0.74 & $6.38\times10^{-1}$ & $1.76\times10^{-3}$ & $9.11\times10^{0}$ & $1.10\times10^{2}$ & $2.66\times10^{-1}$ & $2.42\times10^{0}$ & $4.13\times10^{2}$ & 3.6 & 96.4 \\
\initkp{1} & \finalkp{4} & 0.74 & $2.28\times10^{-1}$ & $3.34\times10^{-3}$ & $2.26\times10^{0}$ & $4.42\times10^{2}$ & $2.66\times10^{-1}$ & $6.02\times10^{-1}$ & $1.66\times10^{3}$ & 50.8 & 49.2 \\
\initkp{1} & \finalkp{5} & 0.73 & $1.28\times10^{-1}$ & $2.42\times10^{-3}$ & $9.17\times10^{-1}$ & $1.09\times10^{3}$ & $2.66\times10^{-1}$ & $2.44\times10^{-1}$ & $4.11\times10^{3}$ & 63.3 & 36.7 \\
\initkp{1} & \finalkp{6} & 0.71 & $5.73\times10^{-2}$ & $2.85\times10^{-3}$ & $8.18\times10^{-1}$ & $1.22\times10^{3}$ & $2.66\times10^{-1}$ & $2.17\times10^{-1}$ & $4.60\times10^{3}$ & 92.3 & 7.7 \\
\initkp{1} & \finalkp{7} & 0.71 & $5.00\times10^{-2}$ & $2.36\times10^{-3}$ & $5.55\times10^{-1}$ & $1.80\times10^{3}$ & $2.66\times10^{-1}$ & $1.47\times10^{-1}$ & $6.78\times10^{3}$ & 91.5 & 8.5 \\
\bottomrule
\end{tabular}
\end{adjustbox}
\end{table}

% Detailed Kramers-pair-resolved PL rates
\begin{table}
\centering
\tiny
\setlength{\tabcolsep}{1.6pt}
\renewcommand{\arraystretch}{0.82}
\caption{Detailed Kramers-pair-resolved PL rates for $V_HEr^{3+}_{Mg2}$. "KP" denotes Kramers pair.}
\label{tab:pl-transition-resolved}
\begin{adjustbox}{max width=\linewidth, max totalheight=0.90\textheight, keepaspectratio}
\begin{tabular}{@{}|l|l|c|c|c|c|c|c|c|c|c|c|@{}}
\toprule
\makecell[c]{Initial\\state} & \makecell[c]{Final\\state} & \makecell[c]{$\Delta E$\\(eV)} & \makecell[c]{$\mu_\mathrm{M}$\\($\mu_\mathrm{B}$)} & \makecell[c]{$\mu_\mathrm{E}$\\(D)} & \makecell[c]{$k_\mathrm{if}$\\(s$^{-1}$)} & \makecell[c]{$\tau_{if}$\\(ms)} & $p_i$ & \makecell[c]{$ k_\mathrm{if}^{eff, Kramers}$\\(s$^{-1}$)} & \makecell[c]{$\tau^{eff, Kramers}$\\(ms)} & \makecell[c]{E\\(\%)} & \makecell[c]{M\\(\%)} \\
\midrule
\initkp{1} & \finalkp{1} & 0.75 & $8.23\times10^{-1}$ & $3.11\times10^{-3}$ & $1.64\times10^{1}$ & 61.138 & $2.56\times10^{-1}$ & $4.19\times10^{0}$ & $2.39\times10^{2}$ & 6.5 & 93.5 \\
\initkp{1} & \finalkp{2} & 0.75 & $4.36\times10^{-1}$ & $1.50\times10^{-3}$ & $4.48\times10^{0}$ & $2.23\times10^{2}$ & $2.56\times10^{-1}$ & $1.15\times10^{0}$ & $8.71\times10^{2}$ & 5.4 & 94.6 \\
\initkp{1} & \finalkp{3} & 0.74 & $6.75\times10^{-1}$ & $1.52\times10^{-3}$ & $1.01\times10^{1}$ & 98.691 & $2.56\times10^{-1}$ & $2.60\times10^{0}$ & $3.85\times10^{2}$ & 2.4 & 97.6 \\
\initkp{1} & \finalkp{4} & 0.74 & $2.14\times10^{-1}$ & $2.98\times10^{-3}$ & $1.90\times10^{0}$ & $5.27\times10^{2}$ & $2.56\times10^{-1}$ & $4.86\times10^{-1}$ & $2.06\times10^{3}$ & 48.4 & 51.6 \\
\initkp{1} & \finalkp{5} & 0.73 & $1.52\times10^{-1}$ & $2.10\times10^{-3}$ & $9.15\times10^{-1}$ & $1.09\times10^{3}$ & $2.56\times10^{-1}$ & $2.35\times10^{-1}$ & $4.26\times10^{3}$ & 48.0 & 52.0 \\
\initkp{1} & \finalkp{6} & 0.72 & $5.54\times10^{-2}$ & $2.19\times10^{-3}$ & $5.06\times10^{-1}$ & $1.98\times10^{3}$ & $2.56\times10^{-1}$ & $1.30\times10^{-1}$ & $7.71\times10^{3}$ & 88.3 & 11.7 \\
\initkp{1} & \finalkp{7} & 0.71 & $4.40\times10^{-2}$ & $1.86\times10^{-3}$ & $3.53\times10^{-1}$ & $2.83\times10^{3}$ & $2.56\times10^{-1}$ & $9.06\times10^{-2}$ & $1.10\times10^{4}$ & 89.7 & 10.3 \\
\bottomrule
\end{tabular}
\end{adjustbox}
\end{table}

% Detailed Kramers-pair-resolved PL rates
\begin{table}
\centering
\tiny
\setlength{\tabcolsep}{1.6pt}
\renewcommand{\arraystretch}{0.82}
\caption{Detailed Kramers-pair-resolved PL rates for $2V_HEr^{3+}_{Mg1}$. "KP" denotes Kramers pair.}
\label{tab:pl-transition-resolved}
\begin{adjustbox}{max width=\linewidth, max totalheight=0.90\textheight, keepaspectratio}
\begin{tabular}{@{}|l|l|c|c|c|c|c|c|c|c|c|c|@{}}
\toprule
\makecell[c]{Initial\\state} & \makecell[c]{Final\\state} & \makecell[c]{$\Delta E$\\(eV)} & \makecell[c]{$\mu_\mathrm{M}$\\($\mu_\mathrm{B}$)} & \makecell[c]{$\mu_\mathrm{E}$\\(D)} & \makecell[c]{$k_\mathrm{if}$\\(s$^{-1}$)} & \makecell[c]{$\tau_{if}$\\(ms)} & $p_i$ & \makecell[c]{$ k_\mathrm{if}^{eff, Kramers}$\\(s$^{-1}$)} & \makecell[c]{$\tau^{eff, Kramers}$\\(ms)} & \makecell[c]{E\\(\%)} & \makecell[c]{M\\(\%)} \\
\midrule
\initkp{1} & \finalkp{1} & 0.76 & $1.04\times10^{0}$ & $7.34\times10^{-5}$ & $2.47\times10^{1}$ & 40.474 & $3.12\times10^{-1}$ & $7.70\times10^{0}$ & $1.30\times10^{2}$ & 0.0 & 100.0 \\
\initkp{1} & \finalkp{2} & 0.75 & $4.61\times10^{-1}$ & $8.88\times10^{-5}$ & $4.74\times10^{0}$ & $2.11\times10^{2}$ & $3.12\times10^{-1}$ & $1.48\times10^{0}$ & $6.77\times10^{2}$ & 0.0 & 100.0 \\
\initkp{1} & \finalkp{3} & 0.74 & $3.62\times10^{-1}$ & $9.95\times10^{-5}$ & $2.77\times10^{0}$ & $3.61\times10^{2}$ & $3.12\times10^{-1}$ & $8.65\times10^{-1}$ & $1.16\times10^{3}$ & 0.0 & 100.0 \\
\initkp{1} & \finalkp{4} & 0.73 & $1.67\times10^{-1}$ & $3.59\times10^{-5}$ & $5.70\times10^{-1}$ & $1.75\times10^{3}$ & $3.12\times10^{-1}$ & $1.78\times10^{-1}$ & $5.62\times10^{3}$ & 0.0 & 100.0 \\
\initkp{1} & \finalkp{5} & 0.72 & $8.43\times10^{-2}$ & $6.21\times10^{-5}$ & $1.38\times10^{-1}$ & $7.26\times10^{3}$ & $3.12\times10^{-1}$ & $4.29\times10^{-2}$ & $2.33\times10^{4}$ & 0.3 & 99.7 \\
\initkp{1} & \finalkp{6} & 0.71 & $5.98\times10^{-2}$ & $1.02\times10^{-4}$ & $6.89\times10^{-2}$ & $1.45\times10^{4}$ & $3.12\times10^{-1}$ & $2.15\times10^{-2}$ & $4.66\times10^{4}$ & 1.4 & 98.6 \\
\initkp{1} & \finalkp{7} & 0.70 & $7.31\times10^{-2}$ & $6.00\times10^{-5}$ & $9.90\times10^{-2}$ & $1.01\times10^{4}$ & $3.12\times10^{-1}$ & $3.09\times10^{-2}$ & $3.24\times10^{4}$ & 0.3 & 99.7 \\
\bottomrule
\end{tabular}
\end{adjustbox}
\end{table}

% Detailed Kramers-pair-resolved PL rates
\begin{table}
\centering
\tiny
\setlength{\tabcolsep}{1.6pt}
\renewcommand{\arraystretch}{0.82}
\caption{Detailed Kramers-pair-resolved PL rates for $2V_HEr^{3+}_{Mg2}$. "KP" denotes Kramers pair.}
\label{tab:pl-transition-resolved}
\begin{adjustbox}{max width=\linewidth, max totalheight=0.90\textheight, keepaspectratio}
\begin{tabular}{@{}|l|l|c|c|c|c|c|c|c|c|c|c|@{}}
\toprule
\makecell[c]{Initial\\state} & \makecell[c]{Final\\state} & \makecell[c]{$\Delta E$\\(eV)} & \makecell[c]{$\mu_\mathrm{M}$\\($\mu_\mathrm{B}$)} & \makecell[c]{$\mu_\mathrm{E}$\\(D)} & \makecell[c]{$k_\mathrm{if}$\\(s$^{-1}$)} & \makecell[c]{$\tau_{if}$\\(ms)} & $p_i$ & \makecell[c]{$ k_\mathrm{if}^{eff, Kramers}$\\(s$^{-1}$)} & \makecell[c]{$\tau^{eff, Kramers}$\\(ms)} & \makecell[c]{E\\(\%)} & \makecell[c]{M\\(\%)} \\
\midrule
\initkp{1} & \finalkp{1} & 0.75 & $1.05\times10^{0}$ & $2.44\times10^{-3}$ & $2.58\times10^{1}$ & 38.816 & $2.70\times10^{-1}$ & $6.97\times10^{0}$ & $1.44\times10^{2}$ & 2.5 & 97.5 \\
\initkp{1} & \finalkp{2} & 0.75 & $3.38\times10^{-1}$ & $2.13\times10^{-3}$ & $3.03\times10^{0}$ & $3.30\times10^{2}$ & $2.70\times10^{-1}$ & $8.20\times10^{-1}$ & $1.22\times10^{3}$ & 16.1 & 83.9 \\
\initkp{1} & \finalkp{3} & 0.74 & $3.53\times10^{-1}$ & $1.05\times10^{-3}$ & $2.81\times10^{0}$ & $3.56\times10^{2}$ & $2.70\times10^{-1}$ & $7.61\times10^{-1}$ & $1.31\times10^{3}$ & 4.1 & 95.9 \\
\initkp{1} & \finalkp{4} & 0.73 & $2.38\times10^{-1}$ & $1.05\times10^{-3}$ & $1.29\times10^{0}$ & $7.74\times10^{2}$ & $2.70\times10^{-1}$ & $3.49\times10^{-1}$ & $2.86\times10^{3}$ & 8.6 & 91.4 \\
\initkp{1} & \finalkp{5} & 0.73 & $1.50\times10^{-1}$ & $1.62\times10^{-3}$ & $7.25\times10^{-1}$ & $1.38\times10^{3}$ & $2.70\times10^{-1}$ & $1.96\times10^{-1}$ & $5.10\times10^{3}$ & 36.1 & 63.9 \\
\initkp{1} & \finalkp{6} & 0.72 & $7.13\times10^{-2}$ & $1.74\times10^{-3}$ & $3.88\times10^{-1}$ & $2.58\times10^{3}$ & $2.70\times10^{-1}$ & $1.05\times10^{-1}$ & $9.54\times10^{3}$ & 74.2 & 25.8 \\
\initkp{1} & \finalkp{7} & 0.69 & $3.09\times10^{-2}$ & $1.30\times10^{-3}$ & $1.60\times10^{-1}$ & $6.25\times10^{3}$ & $2.70\times10^{-1}$ & $4.33\times10^{-2}$ & $2.31\times10^{4}$ & 89.6 & 10.4 \\
\bottomrule
\end{tabular}
\end{adjustbox}
\end{table}

\bibliography{references}